\documentclass[useAMS,usenatbib]{mn2e}
\usepackage{amssymb,graphicx,amsmath}
\usepackage{color, soul}	
\usepackage{todonotes}	
\usepackage{verbatim}	
\setstcolor{red}
\usepackage[pass,a4paper]{geometry}	

\def\apropto{%
  \def\p{%
    \setbox0=\vbox{\hbox{$\propto$}}%
    \ht0=0.6ex \box0 }%
  \def\s{%
    \vbox{\hbox{$\sim$}}%
  }%
  \mathrel{\raisebox{0.7ex}{%
      \mbox{$\underset{\s}{\p}$}%
    }}%
}

\title[On the Efficiency of Jet Production in AGNs]{On The Efficiency of Jet Production in Radio Galaxies} 
\author[Nemmen]{Rodrigo S. Nemmen,$^{1,2,3,4}$\thanks{E-mail: rodrigo.nemmen@iag.usp.br} Alexander Tchekhovskoy$^{5,6,7}$ \\
$^{1}$Instituto de Astronomia, Geof\'{\i}sica e Ci\^encias Atmosf\'ericas, Universidade de S\~ao Paulo, S\~ao Paulo, SP 05508-090, Brazil \\
$^{2}$NASA Goddard Space Flight Center, Greenbelt, MD 20771, USA \\
$^{3}$Center for Research and Exploration in Space Science \& Technology (CRESST) \\
$^{4}$Department of Physics, University of Maryland, Baltimore County, 1000 Hilltop Circle, Baltimore, MD 21250, USA \\
$^5$Lawrence Berkeley National Laboratory, 1 Cyclotron Road, Berkeley, CA 94720, USA \\$^6$Department of Astronomy and Theoretical Astrophysics Center, University of California, Berkeley, CA 94720-3411, USA \\
$^7$NASA Einstein Fellow \\
}

\begin{document}

\date{Accepted 2015 February 5. Received 2015 January 27; in original form 2014 June 27}

\pagerange{\pageref{firstpage}--\pageref{lastpage}} \pubyear{2014}

\maketitle

\label{firstpage}

\begin{abstract}
The mechanisms that produce and power relativistic jets are fundamental open questions in black hole (BH) astrophysics. In order to constrain these mechanisms, we analyze the energy efficiency of jet production $\eta$ based on archival \emph{Chandra} observations of 27 nearby, low-luminosity active galactic nuclei. We obtain $\eta$ as the ratio of the jet power, inferred from the energetics of jet powered X-ray emitting cavities, to the BH mass accretion rate $\dot{M}_{\rm BH}$. The standard assumption in estimating $\dot{M}_{\rm BH}$ is that all the gas from the Bondi radius $r_B$ makes it down to the BH. It is now clear, however, that only a small fraction of the gas reaches the hole. To account for this effect, 
we use the standard disk mass-loss scaling, $\dot{M}(r) \propto (r/r_{\rm B})^s \dot M_{\rm Bondi}$. 
This leads to much lower values of $\dot M_{\rm BH}$ and higher values of $\eta$ than in previous studies. If hot accretion flows are characterized by $0.5 \leq s \leq 0.6$ -- on the lower end of recent theoretical and observational studies -- then dynamically-important magnetic fields near rapidly spinning BHs are necessary to account for the high $\eta \approx 100-300$ per cent in the sample. Moreover, values of $s>0.6$ are essentially ruled out, or there would be insufficient energy to power the jets. 
We discuss the implications of our results for the distribution of massive BH spins and the possible impact of a significant extra cold gas supply on our estimates.
\end{abstract}


\section{Introduction}

Collimated jets are often observed emanating from astrophysical compact objects accreting magnetized plasma, spanning a huge range of masses and diverse environments: black hole binaries, active galactic nuclei (AGNs), gamma-ray bursts, neutron stars and young stellar objects \citep{Belloni10}. Out of these classes of objects, the most relativistic outflows are produced by the black hole engines in microquasars, radio galaxies and gamma-ray bursts. Despite a wealth of data and abundance of theoretical models, the exact details behind the production of these relativistic jets remain elusive. 

One of the most popular ideas for the powering of relativistic jets is the Blandford-Znajek (BZ) mechanism: the idea that the free energy associated with black hole spin can be tapped by large scale magnetic field lines threading the horizon and carried away from the black hole in an electromagnetic jet \citep{Blandford77,Meier12}. The basic tenets of this model have been confirmed in numerical simulations (e.g., \citealt{Semenov04,2005MNRAS.359..801K,Sasha12rev}). 
There is circumstantial evidence supporting the BZ model in observations of stellar mass black holes, through a correlation between the 5 GHz radio luminosity of the transient ballistic jet launched in the near-Eddington accretion state \citep{Fender12} -- which is a proxy of the jet power -- and the spin parameter estimated via continuum fitting (\citealt{Narayan12,Steiner13}; but see \citealt{Fender10,Russell13bhb}). 

Observational clues for the nature of jet production in active galactic nuclei (AGN) are less clear. Even though spin has been invoked to explain the radio-loudness distribution of AGNs (e.g. \citealt{Sikora07,Sasha10}) as well as various other observations (e.g., \citealt{Gardner14}), the actual estimates of supermassive black hole (SMBH) spins in radio-loud AGN -- crucial for probing BZ mechanisms -- are currently affected by considerable uncertainties and/or are strongly model-dependent (e.g., \citealt{Nemmen07,Martinez-Sansigre11,King13spin,Daly14}). Given the small internal scatter in the correlation between the optical luminosity from the accretion disk and the lobe radio luminosity of FR II radio galaxies, \cite{van-Velzen13} concluded that either these quasars have very similar (high) spins or, alternatively, jet spin powering is not relevant at all. In the context of recent evidence for near-maximum spin of many jet-less SMBHs \citep{Reynolds13}, this result might actually suggest the opposite: that jet-producing SMBHs are also powered by near-maximally spinning BHs.

One promising approach for probing the physical conditions behind jet launching in AGNs is to evaluate the energy efficiency of jet production which we define as $\eta_{\rm jet} \equiv P_{\rm jet}/(\dot{M}_{\rm BH} c^2)$, given appropriate measurements of the jet power $P_{\rm jet}$ \emph{and} the mass accretion rate onto the black hole $\dot{M}_{\rm BH}$. Estimating the efficiency of jet production gives us an indirect diagnostic of the jet formation process and the source of jet power. For instance, high efficiencies would suggest spin powering and possibly dynamically important magnetic fields near the horizon (e.g., \citealt{Sasha11,Zamaninasab14}) as opposed to an outflow from an accretion disk \citep{Blandford82,Ghosh97,Livio99}. $\eta_{\rm jet}$ is also relevant for our understanding of the ``radio mode'' of AGN feedback in massive early-type galaxies (e.g., \citealt{Allen06,McNamara07,Fabian12}) and it impacts the cosmological evolution and growth of massive black holes (e.g., \citealt{Fanidakis11,Volonteri12}).

One of the most straightforward methods of estimating the jet power for radio-loud AGNs is by using the cavities observed in \emph{Chandra} images of the environment of the galaxy as calorimeters of the jet; the jet power is then the energy required to create the cavity divided by the time required to inflate it (e.g., \citealt{Dunn04,Birzan04,Dunn05,Allen06,McNamara12}). In the case of SMBHs, a reliable and clean way of estimating $\dot{M}_{\rm BH}$ is by using \emph{Chandra} observations that probe the temperatures and densities of the plasma at the sphere of influence of the black hole (i.e. the Bondi radius $r_B$). Then, using the simple Bondi model \citep{Bondi52} -- which ignores angular momentum, viscosity, etc -- the inflow rate at the Bondi radius $\dot{M}_B$ can be estimated (e.g., \citealt{Di-Matteo03}). 

Various works estimated $P_{\rm jet}/(\dot{M}_B c^2)$ using the methods outlined above. For instance, \cite{Allen06} observed that $P_{\rm jet} \sim$ a few per cent $\dot{M}_B c^2$ (cf. also \citealt{Merloni07,Balmaverde08,Vattakunnel10,Russell13}). \cite{McNamara11} found that $P_{\rm jet} > \dot{M}_B c^2$ for the most powerful radio galaxies, i.e. Bondi accretion is generally unable to provide enough power to the jets in these systems (cf. also \citealt{Rafferty06}). It should be noted that most of the systems analyzed by McNamara et al. are quite distant, such that large extrapolations are required in order to derive the Bondi rates leading in many cases to several orders of magnitude uncertainty in the Bondi rates.

Therefore, while the Bondi rate is simple and straightforward, it is a crude estimator of the mass accretion rate onto the black hole. The simple Bondi model neglects important phenomena in the accretion flow, most importantly rotation and viscosity/turbulence in the plasma. Important advances in our understanding of accretion theory have been made in the last decade. For instance, it was realized that not all gas fed at the Bondi radius may necessarily accrete onto black holes fed at low rates \citep{Narayan94,Blandford99}. In fact, recent numerical simulations (e.g., \citealt{Yuan12,Yuan12b,Sadowski13sim,Li13}), analytical models (e.g., \citealt{Begelman12}) and X-ray observations of the two low-luminosity AGNs for which we can currently resolve the Bondi radius with \emph{Chandra} -- Sgr A* and NGC 3115 \citep{Wang13,Wong14} -- indicate that in actuality the majority of the gas is lost in-between the Bondi radius and the event horizon in radiatively inefficient accretion flows (RIAFs; cf.   \citealt{Yuan14} for a review), presumably in the form of winds. Nevertheless, previous attempts at estimating the efficiency of jet production using ``X-ray cavity calorimetry'' and Bondi rates have neglected these important advances in accretion theory.

In this work, we set to estimate the efficiency of jet production from SMBHs using current constraints on jet power and inflow rate at the Bondi radius, as estimated from X-ray observations. We collate all available information on these observations and construct a sample of SMBHs accreting at sub-Eddington levels in radio galaxies. We then make use of the observationally-motivated advances in the understanding of RIAFs -- taking into account particularly the issue of mass-loss -- in order to relate the outer boundary conditions at $r_B$ to the inner boundary conditions near the event horizon and re-evaluate jet kinetic efficiencies. Our improved estimates place important constraints on the conditions of jet formation in SMBHs. 

Our paper is organized as follows. In Section \ref{sec:obs} we describe the sample of 27 radio galaxies that we used and the data. Section \ref{sec:corr} evaluates the updated constraints on the relation between jet power and Bondi inflow rate implied by our sample. In Section \ref{sec:eff} we describe the new estimates of the efficiency of jet production, taking into account RIAF models. We discuss the implications of our results as well as the caveats involved in \textsection \ref{sec:disc}. We conclude by presenting a summary of our results in \textsection \ref{sec:end}.

\section{Observations}	\label{sec:obs}

We searched the literature for \emph{Chandra}-observed radio galaxies that show evidence for jet-inflated X-ray cavities in their central regions. These cavities can be used as calorimeters to estimate the jet power. We concentrate on those galaxies that in addition have measurements of gas temperature and density profiles and use them to estimate the Bondi inflow rate due to hot diffuse gas. We find about 53 X-ray luminous radio galaxies for which both estimates of jet power and Bondi rates are available \citep{Allen06, Rafferty06, Balmaverde08, Russell13}.

The jet powers were calculated in two ways. In the first method, the jet powers were estimated as $P_{\rm jet} = E_{\rm cav}/t_{\rm age}$, where $E_{\rm cav}$ is the energy required to create the observed cavities and $t_{\rm age}$ is the age of the cavity (e.g., \citealt{Churazov02,Birzan04,Dunn04,Dunn05,Allen06,Russell13}). The usual assumption in deriving $E_{\rm cav}$ is that the cavities are inflated slowly such that $E_{\rm cav}=4PV$ where $P$ is the thermal pressure of the surrounding X-ray emitting gas, $V$ is the volume of the cavity and the cavity is assumed to be filled up with relativistic plasma. The age of the cavity is usually assumed to be either the sound-crossing timescale $t_{\rm c_s}=D/c_s$ where $D$ is the distance of the bubble centre from the black hole and $c_s$ is the adiabatic sound speed, or the buoyancy timescale $t_{\rm buoy}=D/v_t$ where $v_t$ is the bubble terminal velocity. In the second method, the jet power was estimated from the radio core luminosity \citep{Heinz07,Balmaverde08}.

We estimated the the rate of gas inflowing towards the black hole at large distances in the framework of the Bondi model which is the most simple configuration describing the accretion of gas onto a central black hole \citep{Bondi52}: the model assumes a spherically symmetric flow with negligible angular momentum. The resulting mass inflow rate -- often referred to as the Bondi accretion rate -- can be written as $\dot{M}_{\rm B}=\pi \lambda c_s \rho r^2_B$, where $r_B=2GM/c_s^2$ is the accretion radius (or Bondi radius), $G$ is the gravitational constant, $M$ is the black hole mass, $c_s$ is the sound speed of the gas at $r_A$, $\rho$ is the density of gas at $r_A$ and $\lambda$ is a numerical coefficient that depends on the adiabatic index of the gas. This estimate is frequently used in studies of the central X-ray emitting gas in galaxies (e.g., \citealt{Wang13,Wong14}). The values of $\dot{M}_{\rm B}$ for the systems were calculated from the  gas temperature and density profiles measured with \emph{Chandra}. The Bondi radius is not resolved in our sample and the appropriate density and temperature are determined by extrapolating the observed data, typically by a factor of 3
or greater, in radius (e.g., \citealt{Russell13}). 

Many of the systems we found in the literature are located at considerable distances with the \emph{Chandra} spatial resolution $r_{\rm min}$ considerably exceeding the accretion radius; as such these distant galaxies require a considerable extrapolation into the Bondi sphere making their Bondi accretion rates highly uncertain. In order to select only the sources with the most accurate estimates of $\dot{M}_B$, we adopt the selection criterion in which we only keep galaxies with $r_{\rm min}/r_B<30$. 
As a result of applying our selection criterion, most of the sample of \cite{Rafferty06} is excluded from our analysis because it consists mostly of distant sources\footnote{In the notation of \cite{Rafferty06}, $r_{\rm min}$ corresponds approximately to their semi-major axis of the central region $a$ used for extraction.}. 

Our final sample contains 27 radio galaxies hosting low-luminosity AGNs spanning 3 orders of magnitude in accretion and jet power. Table \ref{tab:sample} displays the galaxies, jet powers and Bondi rates, parametrized as Bondi powers ($P_B \equiv 0.1 \dot{M}_{\rm B} c^2$) as traditionally defined in the literature. The references from which we took the jet powers and Bondi rates are also listed. 

\begin{table}
\centering
\caption{Sample of galaxies, Bondi accretion rates ($P_B \equiv 0.1 \dot{M}_{\rm B} c^2$) and jet powers.} 
\begin{tabular}{@{}lcccccc@{}}
\hline
Galaxy & $\log P_{\rm B}$ & $\log P_{\rm jet}$$^{\rm i}$ & Ref.$^{\rm ii}$ \\
& (erg s$^{-1}$) & (erg s$^{-1}$) & \\
\hline
3C066B & $ 44.59 ^{+ 0.16 }_{- 0.12 } $ & $ 44.25 \pm 0.40 $ & b \\ 
3C083.1 & $ 45.11 ^{+ 0.13 }_{- 0.10 } $ & $ 43.69 \pm 0.40 $ & b \\ 
3C270 & $ 44.41 ^{+ 0.16 }_{- 0.12 } $ & $ 43.77 \pm 0.40 $ & b \\ 
3C296 & $ 45.36 ^{+ 0.20 }_{- 0.14 } $ & $ 44.05 \pm 0.40 $ & b \\ 
3C449 & $ 44.48 ^{+ 0.21 }_{- 0.14 } $ & $ 43.66 \pm 0.40 $ & b \\ 
3C465 & $ 45.15 ^{+ 0.12 }_{- 0.10 } $ & $ 44.58 \pm 0.40 $ & b \\ 
UGC6297 & $ 42.68 ^{+ 0.31 }_{- 0.18 } $ & $ 41.83 \pm 0.40 $ & b \\ 
UGC7386 & $ 43.95 ^{+ 0.24 }_{- 0.16 } $ & $ 43.18 \pm 0.40 $ & b \\ 
UGC7898 & $ 43.81 ^{+ 0.12 }_{- 0.09 } $ & $ 42.53 \pm 0.40 $ & b \\ 
UGC8745 & $ 44.62 ^{+ 0.26 }_{- 0.16 } $ & $ 42.78 \pm 0.40 $ & b \\ 
NGC1399 & $ 44.45 ^{+ 0.14 }_{- 0.11 } $ & $ 42.34 \pm 0.40 $ & b \\ 
NGC3557 & $ 44.02 ^{+ 0.23 }_{- 0.15 } $ & $ 42.87 \pm 0.40 $ & b \\ 
IC1459 & $ 44.48 ^{+ 0.12 }_{- 0.09 } $ & $ 43.89 \pm 0.40 $ & b \\ 
IC4296 & $ 45.23 ^{+ 0.10 }_{- 0.08 } $ & $ 43.94 \pm 0.40 $ & b \\ 
A2199 & $ 43.60 ^{+ 0.30 }_{- 0.18 } $ & $ 42.95 ^{+ 0.11 }_{- 0.16 } $ & r \\ 
Cen A & $ 42.71 ^{+ 0.05 }_{- 0.05 } $ & $ 43.15 ^{+ 0.15 }_{- 0.18 } $ & r \\ 
HCG62 & $ 43.70 ^{+ 0.22 }_{- 0.15 } $ & $ 42.78 ^{+ 0.18 }_{- 0.22 } $ & r \\ 
M84 & $ 44.11 ^{+ 0.03 }_{- 0.03 } $ & $ 42.04 ^{+ 0.20 }_{- 0.26 } $ & r \\ 
M87 & $ 45.30 ^{+ 0.10 }_{- 0.08 } $ & $ 42.90 ^{+ 0.20 }_{- 0.27 } $ & r \\ 
M89 & $ 43.70 ^{+ 0.70 }_{- 0.26 } $ & $ 41.48 ^{+ 0.18 }_{- 0.22 } $ & r \\ 
NGC507 & $ 44.30 ^{+ 0.30 }_{- 0.18 } $ & $ 43.28 ^{+ 0.20 }_{- 0.24 } $ & r \\ 
NGC1316 & $ 43.70 ^{+ 0.70 }_{- 0.26 } $ & $ 41.95 ^{+ 0.26 }_{- 0.28 } $ & r \\ 
NGC4472 & $ 43.95 ^{+ 0.18 }_{- 0.12 } $ & $ 41.85 ^{+ 0.24 }_{- 0.23 } $ & r \\ 
NGC4636 & $ 42.70 ^{+ 0.10 }_{- 0.08 } $ & $ 41.43 ^{+ 0.18 }_{- 0.19 } $ & r \\ 
NGC5044 & $ 43.30 ^{+ 0.30 }_{- 0.18 } $ & $ 42.11 ^{+ 0.16 }_{- 0.21 } $ & r \\ 
NGC5813 & $ 43.00 ^{+ 0.22 }_{- 0.15 } $ & $ 41.85 ^{+ 0.24 }_{- 0.27 } $ & r \\ 
NGC5846 & $ 43.11 ^{+ 0.64 }_{- 0.25 } $ & $ 42.20 ^{+ 0.20 }_{- 0.24 } $ & r \\ 
\hline
\end{tabular}
\begin{flushleft}
\footnotesize 
REFERENCES: b: \cite{Balmaverde08}; r: \cite{Russell13}. \\
NOTES: (i) For the radio galaxies from \cite{Balmaverde08}, $P_{\rm jet}$ was estimated using the radio core luminosities; for the ones from \cite{Russell13}, it was estimated from X-ray cavities' calorimetry as described in section \ref{sec:obs}.
(ii) References with the additional data needed for estimating $P_{\rm jet}$ for the Russell et al. sample ($PV$, $t_{\rm age}$ etc) are listed in Table 1 of \cite{Russell13}. 
\end{flushleft}
\label{tab:sample}
\end{table}

There is a some overlap between the samples studied by \cite{Rafferty06}, \cite{Allen06}, \cite{Merloni07} and \cite{Russell13}. Whenever there is overlap, we choose the observations of \cite{Russell13} since they correspond to an improved analysis of the nine sources previously studied by \cite{Allen06} with new, deeper \emph{Chandra} exposures. 

We note that \cite{Vattakunnel10} provide estimates of jet power and $\dot{M}_B$ for two additional radio-loud AGNs which do not display extended radio jets or X-ray cavities. For this reason, their jet power estimates are based on an indirect method that relies on using a correlation between core radio luminosity and jet power. In our sample, we only keep sources for which $P_{\rm jet}$ was estimated directly from X-ray cavities used as jet calorimeters and hence we exclude the two AGNs studied by \cite{Vattakunnel10}. 

In the analysis that follows, we accept the estimates of the previously estimated Bondi accretion rates and jet powers at face value. In section \ref{sec:bondi} we discuss the uncertainties involved in such estimates.

\section{An update on the relation between Bondi rates and jet powers}	\label{sec:corr}

\begin{figure}
\centering
\includegraphics[width=\linewidth,trim=60 0 100 10,clip=true]{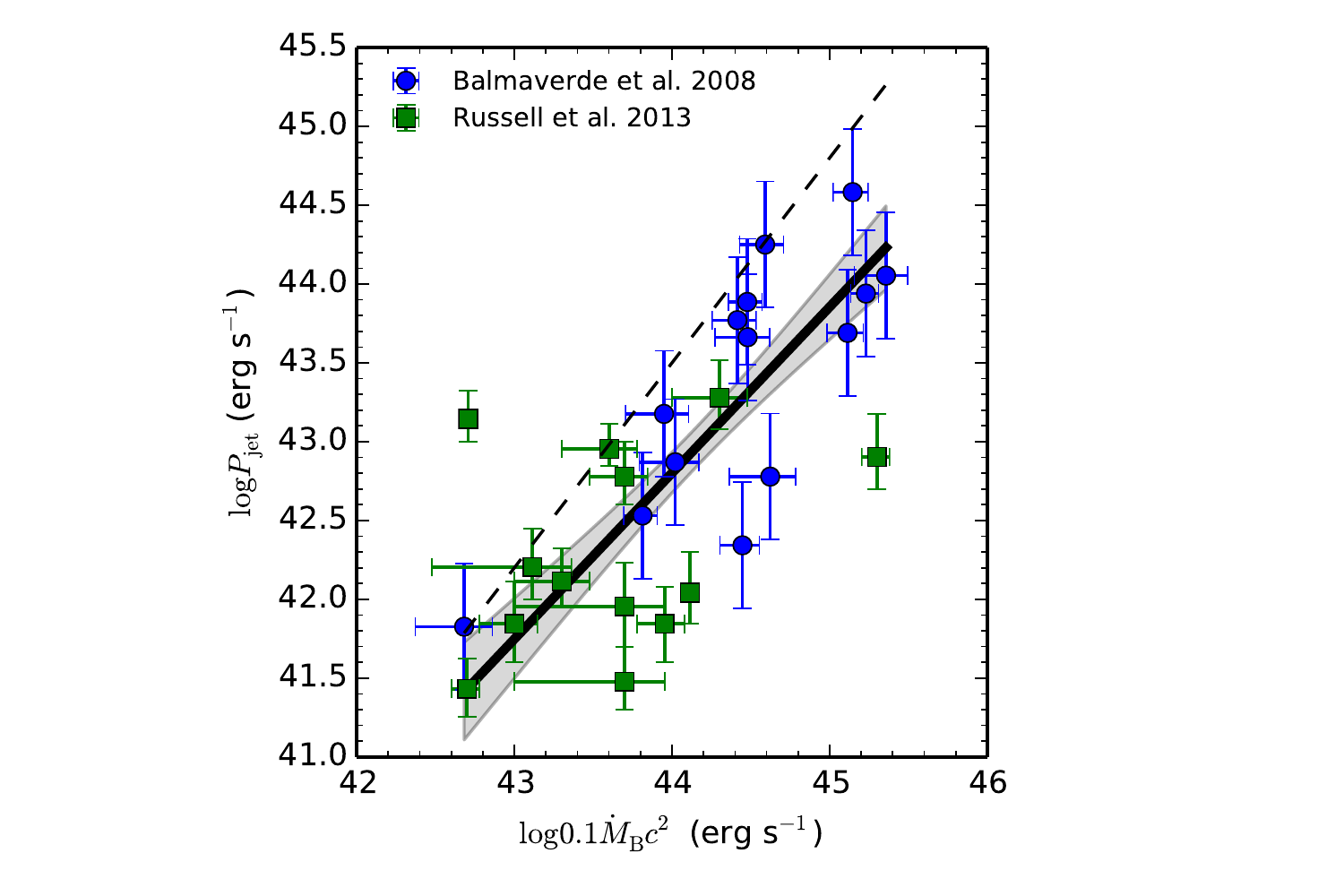}
\caption{Relation between the ``Bondi power'' ($0.1 \dot{M}_{\rm B} c^2$) and jet power for our sample. The thick solid line corresponds to our power-law best-fit to the data (shaded regions corresponds to the $1\sigma$ confidence band of the fit) whereas the thin dashed line corresponds to the best-fit relation obtained by \citet{Allen06}. }
\label{fig:powers}
\end{figure}

Figure \ref{fig:powers} displays the relation between the Bondi and jet powers for our sample, which can be compared with previous analysis based on smaller samples (e.g., \citealt{Allen06}). 
We perform a power-law fit to the $P_{\rm jet}-P_{\rm B}$ data parametrized as a linear fit. We adopt the BCES $Y|X$ regression method (bivariate correlated errors and intrinsic scatter) \citep{Akritas96bces} which takes into account measurement errors in both the ``$X$'' and ``$Y$'' coordinates and the intrinsic scatter in the data\footnote{Uncertainties on the parameters derived from the fits are estimated after carrying out 10000 bootstrap resamples of the data.}. This method has been widely used in fitting datasets in the astronomical community (e.g., \citealt{Allen06,Nemmen12,Sifon13}). 

We obtain a best-fit model $\log P_{\rm jet} = A \log P_{\rm B} + B$ where $A=1.05 \pm 0.19$ and $B=-3.51 \pm 8.49$. Hence the correlation is consistent with a linear correlation within $1\sigma$. The total scatter about the best-fit relation 
is $\approx 0.7$ dex. 
Figure \ref{fig:powers} compares our best-fit model with the previous result of \cite{Allen06} and illustrates that the Allen et al. relation is substantially different from the present one: for a given value of $\dot{M}_{\rm B}$, the mean jet powers predicted by the Allen et al. relation are larger than the ones predicted by our fit by $\approx 0.5$ dex on average. We find a significantly larger scatter in the $P_{\rm B}-P_{\rm jet}$ correlation compared to the 0.1 dex scatter inferred by Allen et al. Therefore, the new relation between accretion rates and jet powers is considerably less tight than previously claimed, a result that was already noticed by \cite{Russell13} within the scope of their analysis of 13 radio galaxies. 

On average, $P_{\rm jet} / (\dot{M}_B c^2) \sim 1\%$; in other words, 1\% of the rest mass energy associated with the Bondi inflow rate is converted to jet power, in agreement with previous estimates \citep{Allen06,Merloni07,Balmaverde08}. In the next sections, we will improve on the estimate of the kinetic efficiency, taking into account improvements beyond the Bondi model in our understanding of the fate of gas accreting  at a low-rate onto SMBHs.

\section{Efficiency of jet production}	\label{sec:eff}

Our underlying assumption is that given the boundary conditions provided by the \emph{Chandra} observations of hot inflowing ISM, we can estimate the accretion rate onto the SMBH as some function $\dot{M}_{\rm BH} = \dot{M}_{\rm BH}(\dot{M}_B)$. As such, our estimates of the kinetic efficiencies translate to
\begin{equation}	\label{eq:eff}
\eta_{\rm jet} = \frac{P_{\rm jet}^{\rm obs}}{\dot{M}_{\rm BH}(\dot{M}_B^{\rm obs}) c^2}
\end{equation}
where the superscript ``obs'' explicitly denotes the observationally derived quantities. We reiterate that $\eta_{\rm jet}$ is defined in this way -- in terms of the of the innermost accretion rate -- because it gives the best diagnostic on scenarios for jet production such as the BZ jet powering.

The magnetized gas reaching galactic nuclei will unavoidably have angular momentum which will modify the character of accretion with respect to the simple Bondi model. The theory of accretion flows has advanced considerably in the last two decades and has  established that turbulent viscosity -- driven by magnetic fields -- as well as outflows, radiative cooling and possibly convection, will certainly modify the character of accretion (e.g., \citealt{Narayan08,McKinney12,Li13,Yuan14} and references therein). Therefore, in order to obtain a more physically realistic estimate of the mass accretion rate onto SMBHs in our sample, our task now is to seek improvements over the simple Bondi mode which capture the more complex accretion physics in-between $r_B$ and the Schwarzschild radius $r_S \equiv 2 GM/c^2$.

Given that the radio galaxies in our sample are all known to host highly sub-Eddington, low-luminosity AGNs (LLAGNs; \citealt{Merloni07,Balmaverde08,Russell13}), we adopt the RIAF framework. The RIAF model is widely regarded as the model that best accounts for the properties of low-luminosity AGNs including our Galactic Center (e.g., \citealt{Yuan03,Wu07,Wang13,Nemmen14, Yuan14}). We first estimate a likely upper limit on $\dot{M}_{\rm BH}$ based on the theory of ``classical'' RIAFs, according to which all gas supplied at $r_B$ is accreted onto the black hole and hence provides a lower limit on $\eta_{\rm jet}$. Then, we incorporate in our estimates the current understanding of RIAFs and the likely mass-loss involved in-between $r_B$ and $r_S$.

\subsection{Kinetic efficiencies based on the ADAF model}	\label{sec:adaf}

Assuming that all gas supplied at the outer radius of the RIAF ends up being accreted onto the black hole, the RIAF solution simplifies to the particular case of advection-dominated accretion flows (ADAF; \citealt{Narayan94}) which corresponds to the viscous rotating analog of spherical Bondi accretion. The density profile in ADAFs is given by $\rho \propto r^{-3/2}$ and the corresponding mass accretion rate onto the black hole is related to the Bondi rate as $\dot{M}_{\rm adaf} \sim \alpha  \dot{M}_B$ in the self-similar approximation where $\alpha<1$ is the standard $\alpha$ prescription for the viscosity \citep{Shakura73, Narayan94,Narayan02}. The more detailed calculations of \cite{Narayan11} result in $\dot{M}_{\rm adaf} \approx 0.3 \dot{M}_B$ for $\alpha=0.1$. 

We base our choice of the value of $\alpha$ on magnetohydrodynamic (MHD) numerical simulations of accretion flows. For example, shearing box simulations find $\alpha \sim 0.01-0.003$ \citep{Hawley11}; global simulations find $\alpha \sim 0.05-0.2$ (e.g., \citealt{Hawley01, Penna13}) far away from the black hole ($R>10 GM/c^2$) which is our region of interest. 
Hence, we adopt $\alpha=0.1$ in our calculations to err on the conservative side as widely used in the literature (e.g., \citealt{Yuan14}). 

\begin{figure}
\centering
\includegraphics[width=\linewidth,trim=0 0 200 20,clip=true]{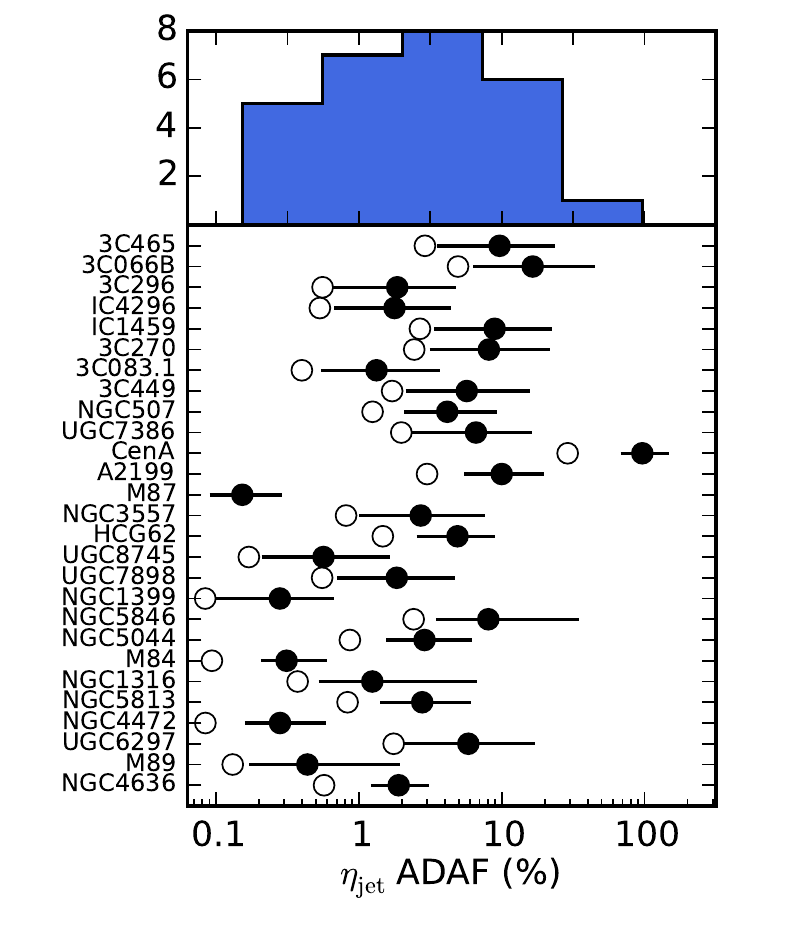}
\caption{Distribution of per cent kinetic efficiencies for the sample using the ADAF model in order to estimate the mass accretion rates. The lower panel shows the individual efficiencies for each radio galaxy; the filled circles correspond to $\alpha=0.1$ (error bars, 1 s.d.) and the empty ones correspond to $\alpha=1$. The upper panel shows the histogram of values of $\eta_{\rm jet}$ for $\alpha=0.1$. }
\label{fig:adaf}
\end{figure}

Figure \ref{fig:adaf} shows the distribution of kinetic efficiency estimates for our sample where we assume the existence of a giant ADAF extending from $r_B$ down to the black hole horizon with $\alpha=0.1$ such that $\dot{M}_{\rm BH} \approx 0.3 \dot{M}_B$. For reference, we also plot in this figure the resulting efficiencies for $\alpha=1$, i.e. $\dot{M}_{\rm BH} \approx \dot{M}_B$ \citep{Narayan11} and the efficiency is simply given as $P_{\rm jet}/(\dot{M}_B c^2)$. The statistical uncertainty on $\eta_{\rm jet}$ was estimated by carrying out a Monte Carlo error propagation based on the uncertainties in $P_{\rm jet}$ and $\dot{M}_B$. The typical uncertainty on $\eta_{\rm jet}$ is $\approx 0.4$ dex (median $1\sigma$ uncertainty).

For $\alpha=0.1$, the median efficiency is $\approx 2.8^{+9.0}_{-2.1}\%$ ($\approx 1\%$ for $\alpha=1$). 
The lowest and highest efficiencies in the sample are achieved by M87 ($\approx 0.1\%$) and the Centaurus cluster ($\approx 90 \%$), respectively. The sample standard deviation corresponds to 0.6 dex. The accretion rates estimated with the ADAF model can be thought of as upper limits to the accretion rate onto the black hole, since the model ignores mass-loss as we will discuss in the next section. Therefore, the kinetic efficiencies estimated in the framework of the ADAF model should be regarded as lower limits on the efficiency of jet production.

\subsection{Kinetic efficiencies based on the ADIOS model}	\label{sec:adios}

In the early analytical studies of RIAFs, \cite{Narayan94,Narayan95b} noted that the positivity of the Bernoulli parameters throughout the flow may lead to the natural production of outflows, therefore introducing potential ``gas leakage'' in ADAFs. \cite{Blandford99} proposed a simple scaling of accretion rate with radius taking into account mass-loss such that  
\begin{equation}	\label{eq:mdot}
\dot{M}(r) = \dot{M}_{\rm o} \left( \frac{r}{r_o} \right)^s,
\end{equation}
where $\dot{M}_{\rm o} = \dot{M}_{\rm adaf} \sim \alpha \dot{M}_{\rm B}$ is the accretion rate at the outer radius $r_o \sim R_{\rm B}$ of the accretion flow and $s$ is a ``mass-loss'' index in the range $0 \leq s \leq 1$. The corresponding density profile is given by 
\begin{equation}	\label{eq:rho}
\rho(r) \propto r^{-\beta}, \qquad  \beta \equiv 3/2 - s
\end{equation}
for $r_S \lesssim r \lesssim r_B$. 
$s=0$ corresponds to a classical ADAF. Models with $s>0$ have been called ADIOS (adiabatic inflow-outflow solution); recently \cite{Begelman12} favored $s=1$ (or $\beta=0.5$). 

The general scaling of equations \ref{eq:mdot} and \ref{eq:rho} has been widely supported by the results of several magnetohydrodynamic numerical simulations of RIAFs (e.g., \citealt{Stone99, Hawley01, Igumenshchev03, Proga03}), in particular the more recent ones (e.g., \citealt{Pang11, McKinney12, Yuan12,Sadowski13sim}). For instance, \cite{Yuan12} find $0.54 \lesssim s \lesssim 0.65$ and $0.65 \lesssim \beta \lesssim 0.85$ for $r>10 r_s$. \cite{McKinney12,Sadowski13sim} find similar results, although they estimate $s$ in the limited radial range $6 r_S \lesssim r \lesssim 15 r_S$ \citep{McKinney12} and $10 r_S \lesssim r \lesssim 100 r_S$ \citep{Sadowski13sim}. \cite{Narayan12grmhd} on the other hand find no strong mass-loss for $r \lesssim 100 r_S$, noting that they are not able to confidently estimate the amount of mass-loss for $r \gtrsim 100 R_S$ (cf. discussion in \citealt{Yuan12}). 

On the observational side, $s$ has been estimated with a variety of methods as ranging between $\approx 0.5$ and $\approx 1$ (e.g., \citealt{Baganoff03,Marrone07,Wang13,Wong11,Wong14,Kuo14}). Particularly, a lower limit on $s$ has been estimated for Sgr A* as $\approx 0.6$ from an upper limit on $\dot{M}_{\rm BH}$ based on sub-millimeter Faraday rotation measurements \citep{Marrone07} and a previous \emph{Chandra} X-ray spectrum observation \citep{Baganoff03}; $s \approx 1$ has been estimated from fitting a more recent 3 Ms \emph{Chandra} observation of the quiescent X-ray spectrum \citep{Wang13}). For NGC 3115, fitting the density profile within 3'' (i.e. within $r_B$) measured with \emph{Chandra} results in $s=0.88^{+0.29}_{-0.20}$ or $\beta=0.62^{+0.20}_{-0.29}$; when fitting a wider range of radii (within 5'') \cite{Wong14} find that $s \approx 0.5$ or $\beta \approx 1$ is roughly consistent with their results.
Such high values of the mass-loss parameter suggest that the inflow is essentially balanced by outflowing gas with only a small fraction of inflow rate actually being accreted onto the hole.

Motivated by these theoretical and observational results, we adopt the ADIOS self-similar model in order to estimate $\dot{M}_{\rm BH}$  and $\eta_{\rm jet}$. We adopt $\dot{M}_{\rm BH} \approx \dot{M}(10 r_s)$ where we keep $s$ free in order to explore its effect on the efficiencies. We adopt $\alpha=0.1$ as discussed previously and $r_{\rm B} = 10^5 r_s$ which corresponds to the typical Bondi radius in elliptical galaxies \citep{Di-Matteo03}. 

The median kinetic efficiency computed in section \ref{sec:adaf} with the ADIOS model as a function of $s$ is given by 
\begin{equation}
{\rm median} \ \eta_{\rm jet}=  2.8 \times 10^{4s}  \, {\rm per \ cent},
\end{equation}
where the $10^{4s}$ factor corresponds to the $\left[ r_B/(10 r_S) \right]^s$ ratio. In other words, the median $\eta_{\rm jet}$ depends strongly on $s$. This can be easily understood: the stronger the mass-loss, the less mass ends up being accreted by the black hole. Therefore, for fixed $\dot{M}_B$ and $P_{\rm jet}$, the central engine must have a higher jet production efficiency in order to account for a fixed jet power as $\dot{M}_{\rm BH}$ decreases.

\begin{figure}
\centering
\includegraphics[width=\linewidth]{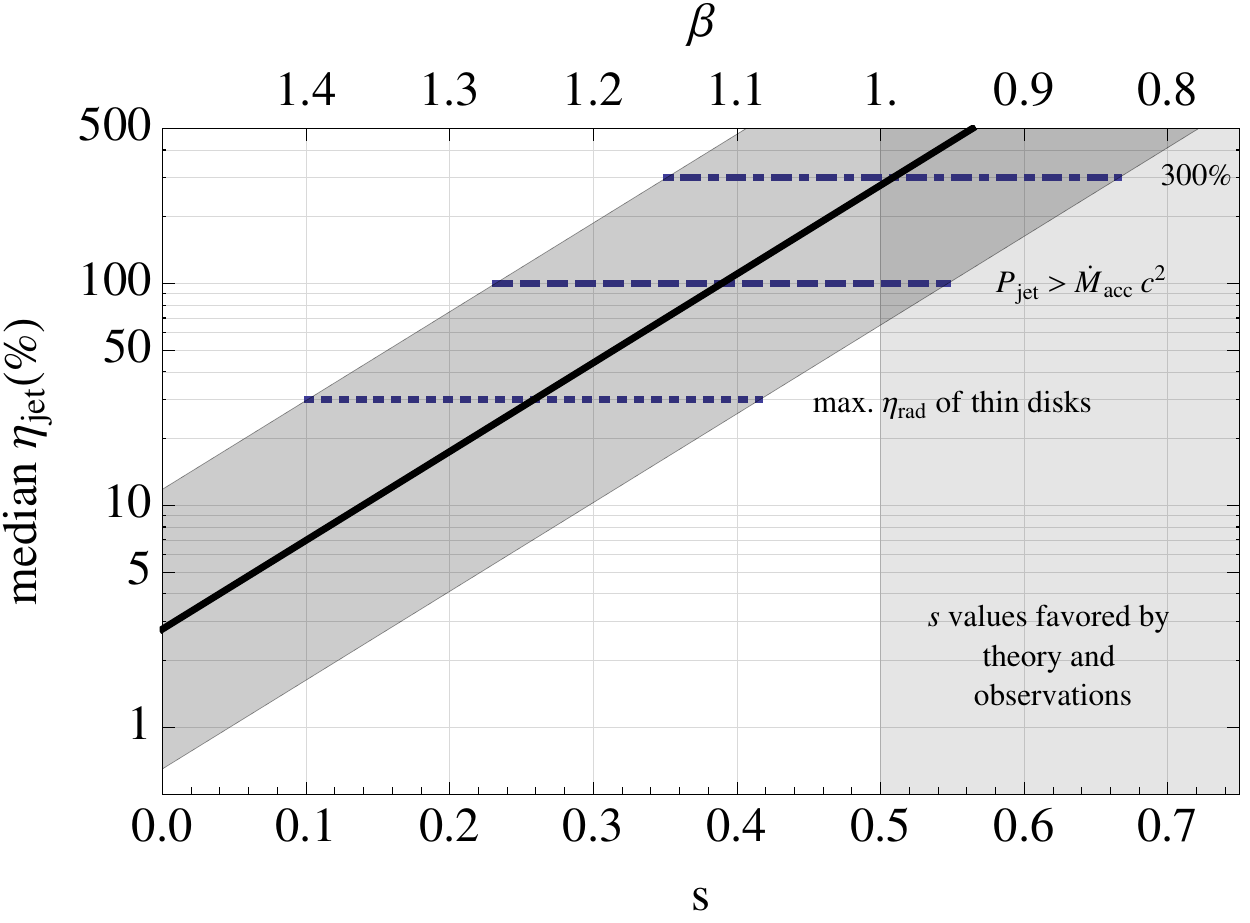}
\caption{Per cent jet kinetic efficiencies as a function of the mass-loss index $s$ for the sample ($\dot{M} \propto r^s$ or equivalently $\rho \propto r^{-\beta}$), where we adopt the ADIOS density profile, $\alpha=0.1$ and $r_B=10^5 r_s$. The solid line represents the median efficiency of the sample and the shaded region around the black solid line corresponds to the scatter around the median. The dashed lines denote the efficiencies 30\%, 100\% and 300\%, respectively. We display the parameter space corresponding to $s>0.5$ as the shaded region to the right, which corresponds to the range of density profiles favored by observations of Sgr A*, NGC 3115 and different theoretical arguments (see text).}
\label{fig:median}
\end{figure}

Figure \ref{fig:median} shows the efficiencies as a function of $s$ up to 500\%. We can see that for $s=0$ we recover the efficiencies estimated with the ADAF model.  In order to appreciate a few notable results in in this plot, let us denote the value of $s$ above which $\eta_{\rm jet}$ exceeds a certain per cent efficiency $\epsilon$ as $s_\epsilon \equiv s(\eta_{\rm jet} > \epsilon)$. We will explore the three critical values $s_{30}$, $s_{100}$ and $s_{300}$ at which the median $\eta_{\rm jet}$ exceeds three relevant efficiencies of energy extraction from accreting black holes.
The maximal radiative efficiency of a thin accretion disk around a black hole with a spin $a=0.998$ is $\approx 30\%$ \citep{Thorne74}; our sample efficiencies exceed 30\% at the relatively low value $s_{30}=0.26 \pm 0.16$\footnote{The error bar denotes the uncertainty in $s$ due to the observed scatter in $P_{\rm jet}/(\dot{M}_B c^2)$.}.
The jet power exceeds the rest mass energy associated with the accreting gas $\dot{M}_{\rm BH} c^2$ (i.e. $P_{\rm jet} > \dot{M}_{\rm BH} c^2$) at $s_{100}=0.39 \pm 0.16$. 
Lastly, for a mass-loss index of $s_{300}=0.51 \pm 0.16$, the kinetic efficiency of the sample exceeds $\approx 300\%$. This is the highest kinetic efficiency achieved so far in current general relativistic magnetohydrodynamic simulations (GRMHD) of jet production from spinning black holes accreting through magnetically arrested flows \citep{Sasha11,McKinney12}. 
It is worth noting that for $s=1$ which is the value recently favored for Sgr A* from fitting the quiescent \emph{Chandra} X-ray spectrum \citep{Wang13} as well as in recent ADIOS models \citep{Begelman12}, the resulting efficiencies considerably exceed 1000\%: there is just not enough accreting rest mass energy or even black hole spin energy via BZ processes to power the jets. We will return to this point in section \ref{sec:disc}.

\begin{figure}
\centering
\includegraphics[width=\linewidth,trim=0 0 200 20,clip=true]{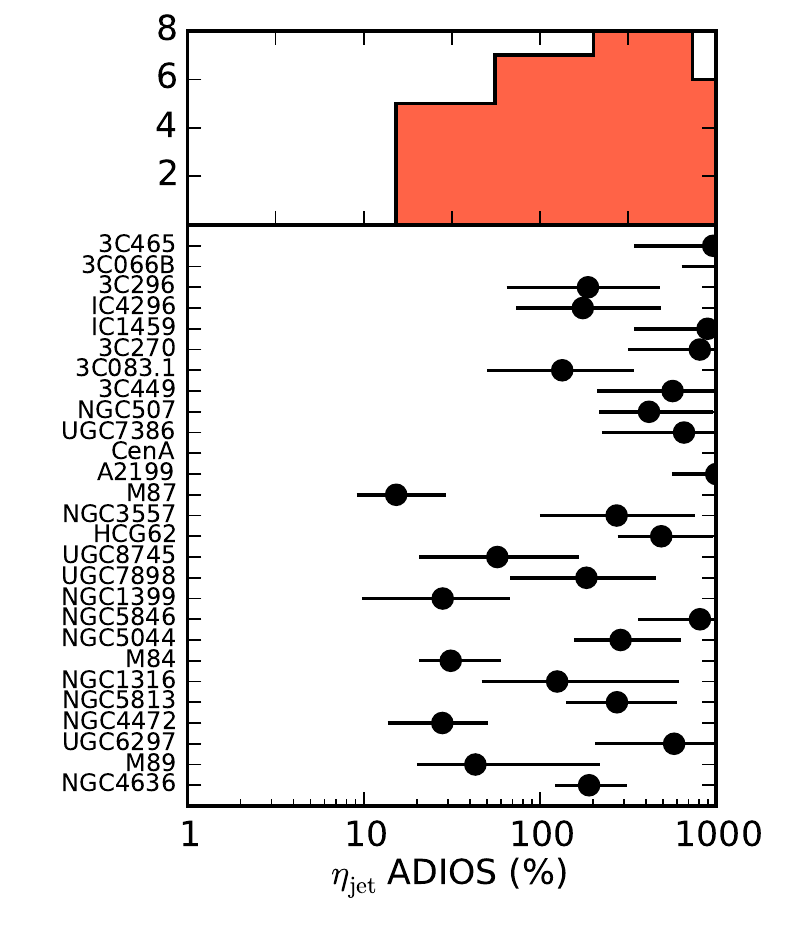}
\caption{Distribution of per cent kinetic efficiencies for the sample using the ADIOS model with $s=0.5$ (or $\beta=1$). The lower panel shows the individual efficiencies for each radio galaxy. The upper panel shows the histogram of values of $\eta_{\rm jet}$. The other parameters are fixed at $\alpha=0.1$ and $r_B=10^5 r_s$. Uncertainties estimated as in Fig. \ref{fig:adaf}.}
\label{fig:adios}
\end{figure}

We show in Figure \ref{fig:adios} the kinetic efficiencies estimated for the individual radio galaxies computed with the ADIOS model adopting $s=0.5$ or $\beta=1$. This value of $\beta$ is broadly consistent with the density profile within 5'' around the Bondi radius for NGC 3115 \citep{Wong14} and is on the lower end of the estimates for Sgr A* \citep{Baganoff03,Marrone07,Wang13} and from recent numerical simulations \citep{Yuan12,McKinney12,Sadowski13sim}. For this value of the mass-loss index -- which corresponds to $\dot{M}_{\rm BH} \sim 10^{-3} \dot{M}_B$ -- there are only four galaxies which have $\eta_{\rm jet}<100\%$ within $1\sigma$: NGC 4472, M84, NGC 1399 and M87. All the other objects have jet powers which can exceed $\dot{M}_{\rm BH} c^2$ given the $1\sigma$ uncertainties; in particular, 11 of them have median efficiencies exceeding 300\%.

\section{Discussion}	\label{sec:disc}

The driving question that we set to attack in this work is to estimate the efficiency of jet production in radio galaxies in terms of the accretion rates at the BH, combining the most up to date observational constraints on the feeding of SMBHs and total jet powers with current constraints on RIAF density profiles. To this end, we compiled from the literature updated data on radio galaxies with both Bondi inflow rates -- which can be used to estimate the hot gas inflow at the radius where the accretion flow begins -- and jet powers estimated from X-ray cavities. Our sample was selected such that the \emph{Chandra} spatial resolution $r_{\rm min}$ is reasonably close to $r_B$ ($r_{\rm min} < 30 r_B$), in order to minimize possible extrapolations in radius which are required to derive the gas density and temperature at $r_B$, needed for computing $\dot{M}_B$. Therefore, our sample consists of relatively nearby radio galaxies.

Our selection criteria of the sample is not unbiased. By virtue of our method of estimating jet powers, we require hot X-ray bright emitting atmospheres displaying cavities which we use as jet calorimeters. For such sources, therefore, (i) jet feedback was efficient in the past, (ii) the SMBHs reside in massive elliptical galaxies with many being cDs and (iii) the bubble-blowing jets are misaligned with our line sight. Because of iii, we are missing blazar-like systems or system with jets but not cavity structures. Due to i and ii, we are selecting AGNs with powerful jets on average -- for which the BZ process is operating efficiently -- and such BHs are on the high-end of the mass distribution, which are expected in the hierarchical framework of galaxy formation to have high spins \citep{Volonteri07}. Such high spins are expected to occur due to the coalescences of two SMBHs that give rise to giant ellipticals (e.g., \citealt{Hughes03}). Therefore, our selection criteria are favoring radio loud AGNs with potentially rapidly spinning SMBHs. 

Our $\dot{M}_B c^2 - P_{\rm jet}$ relation results in jet powers which are a factor of $\approx 3$ lower for a given Bondi rate when compared with the \cite{Allen06} result. In addition, we find a  less tight correlation with a scatter of $\approx 0.6$ dex\footnote{This scatter can be seen as the sample standard deviation in Fig. \ref{fig:adaf}.}. The weaker evidence for the correlation between $\dot{M}_B$ and $P_{\rm jet}$  is in agreement with the results of \cite{Russell13} as expected, given that nearly half of our sample was drawn from their study. 

On physical grounds, a correlation between $\dot{M}_B$ and  $P_{\rm jet}$ is not surprising. The accretion (or viscous) timescale at $r_B$ for an ADAF is
\begin{equation}
t_{\rm vis} \sim 9 \times 10^4 \left( \frac{0.3}{\alpha} \right) \left( \frac{M}{10^9 M_\odot} \right) \left( \frac{r}{10^5 r_s} \right)^{3/2} {\rm years},
\end{equation}
which is considerably shorter than the dynamical timescale of the surrounding medium and the turbulence timescale at $r_B$ \citep{Narayan11}. The accretion flow is therefore expected to be relatively stable and the inflowing gas is expected to be processed into jets and outflows relatively rapidly.

One of the central results of this paper is that given the \emph{measured} jet powers and Bondi rates in our sample of LLAGNs, and taking into account the body of evidence favoring RIAFs for which the mass accretion rate onto the black hole is considerably reduced with respect to the Bondi rate ($\dot{M}_{\rm BH}/\dot{M}_B \ll 1$) -- of which the best examples are currently our Galactic Center and NGC 3115 -- then extremely high efficiencies ($P_{\rm jet} \gtrsim \dot{M}_{\rm BH} c^2$) of jet powering are implied for the bulk of the sample. 

We now discuss the possible implications of these results in the context of current scenarios for jet production around Kerr black holes. Then we critically assess our basic assumptions and explore the caveats and alternative interpretations.

\subsection{The high efficiencies of jet production}

As previously pointed out, the density profiles of RIAFs can be characterized by $\rho(r) \propto r^{-\beta}$ with $0.5 \leq \beta \leq 1$ or equivalently $\dot{M}(r) \propto r^s$ with $0.5 \leq s \leq 1$. 
Taking these constraints on $s$ and $\beta$ at face value for LLAGNs in radio galaxies, we find that the median jet production efficiency of our sample exceeds 300\% for $s \gtrsim 0.5$; even though a few sources at $s=0.5$ would have $\eta_{\rm jet} < 100\%$, 11 sources (41\% of the sample) have efficiencies above 300\%. 

Is it possible at all to obtain such high efficiencies of energy extraction from accreting black holes? There are two viable mechanisms to power jets in galactic nuclei:  (i) by tapping a fraction of the accretion energy and (ii) by confining a strong magnetic field around a spinning black hole and extracting spin energy from the hole via the BZ mechanism. In the case of an accreting Schwarzschild black hole, the energy extraction efficiency is of course limited to the rest mass energy associated with accreting material; if the black hole is surrounded by a magnetically arrested disk (hereafter MAD; \citealt{Bisnovatyi-Kogan74,Narayan03}) the efficiency could be close to 50\% but not much higher. On the other hand, Kerr black holes do not suffer from this limitation. Since about $\sim 30\%$ of the gravitational mass of a maximally rotating black hole is available to be tapped, the central engine could in principle tap the spin energy of the hole with efficiencies $>100\%$. 

Over the last few years, GRMHD simulations have shown that rapidly rotating black holes are able to produce relativistic jets via the BZ mechanism (e.g. \citealt{Komissarov01,Koide02, De-Villiers03, McKinney04,McKinney05,Hawley06,Sasha10,Sasha11,Sasha12rev,Fragile12jet} and references therein). For instance, the simulations of \cite{Sasha11,Sasha12,McKinney12} have demonstrated that $\eta_{\rm jet}>100\%$ can be attained provided that the black hole has a high spin and enough magnetic flux is accumulated near the horizon; i.e. the jet production efficiency through the BZ process is maximized when the black hole is flooded with large-scale magnetic flux and is accreting in the MAD state (cf. also \citealt{Semenov04}). In fact, efficiencies as high as $\approx 350\%$ have been reported \citep{McKinney12,Sasha12rev}. 

If we accept the estimated jet powers and Bondi rates at face value as well as the emerging evidence that RIAFs are subject to considerable mass-loss as in the ADIOS scenario, it follows naturally that most radio galaxies produce jets with powers exceeding the rest mass energy of the material accreted onto the black hole. A likely explanation for such high $\eta_{\rm jet}$ inferred in our analysis is the efficient tapping of the spin energy of SMBHs accreting in the MAD state via BZ processes. In other words, the high values of $\eta_{\rm jet}$ would constitute evidence that powerful, relativistic jets are produced during episodes of accumulation of magnetic flux around rapidly rotating black holes \citep{Sikora13}. 

The interpretation above is supported by the recent results of \cite{Zamaninasab14,Ghisellini14}. \cite{Zamaninasab14} reported a tight correlation between the jet magnetic flux and the accretion disk luminosity for a sample of 76 radio-loud AGNs -- comprised mostly of flat-spectrum radio quasars -- while \cite{Ghisellini14} found that jet powers in blazars exceed the accretion luminosity. The results obtained by Zamaninasab et al. and Ghisellini et al. are best explained if jet-producing SMBHs are accreting via MADs.

An alternative interpretation of our results is possible if the density profiles in radio galaxies are such that most of $\dot{M}_B$ makes its way down to the horizon \citep{Narayan11}. In the framework of the ADIOS model, values of the mass-loss index $s \lesssim 0.4$ would result in the more comfortable efficiencies $<100\%$. For instance, \cite{Narayan12grmhd} find no significant outflows or convection in their GRMHD simulations of RIAFs for $r<100 r_S$. If this is the case, $\eta_{\rm jet}$ would be much lower and easily accommodated by BZ models with moderate levels of magnetic flux \citep{Nemmen07}, without requiring MADs.  
At least for one radio galaxy, M87, the possibility of $s \sim 0$ has been disfavored by recent sub-millimeter observations which allow a constraint on the Faraday rotation measure and hence on the accretion rate: $\dot{M}_{\rm BH} \lesssim 7 \times 10^{-3} \dot{M}_B$ \citep{Kuo14}\footnote{More precisely, \cite{Kuo14} estimate $\dot{M}$ at $21 r_S$ which for practical purposes is approximately the same as our definition of $\dot{M}_{\rm BH} \equiv \dot{M}(r=10 r_S)$.}. In the framework of self-similar ADIOS models, the Kuo et al. constraint on $\dot{M}_{\rm BH}$ suggests that $s > 0.4-0.5$ for M87 (considering $\alpha=0.3-1$). Clearly, progress on the observational side is needed in order to better understand the density profiles in radio galaxies.

One additional possibility is that $\dot{M}_B$ may not be an appropriate indicator of the inflow rates at the outer radius of the accretion flow. For instance, it could be that $\dot{M}_B$  considerably underestimates the true inflow rate due to significant amount of cold gas at $\sim r_B$ and/or the gas inflow at such distances in elliptical galaxies is predominantly cold and chaotic as envisioned by \cite{Gaspari13}. If that is the case, then the amount of gas reaching the black hole could increase significantly, compensating the ADIOS mass-loss. We discuss this possibility in more details in the section below.

\subsection{Constraint on black hole spins and upper limit on RIAF mass-loss index}

One of the advantages of the MAD model is that the inner flow properties are independent of the initial value of magnetic flux fed to the black hole \citep{Sasha12,Sasha12rev}. MAD simulations result in the approximate dependence of jet production efficiency  
\begin{equation}	\label{eq:mad}
\eta_{\rm MAD} \approx 130 \left( \frac{h}{0.3} \right) a^2 \ {\rm per \ cent}
\end{equation}
where $h \equiv H/r$ (\citealt{Sasha12,sasha15}; Tchekhovskoy \& Nemmen, in preparation). 

\begin{figure}
\centering
\includegraphics[width=\linewidth]{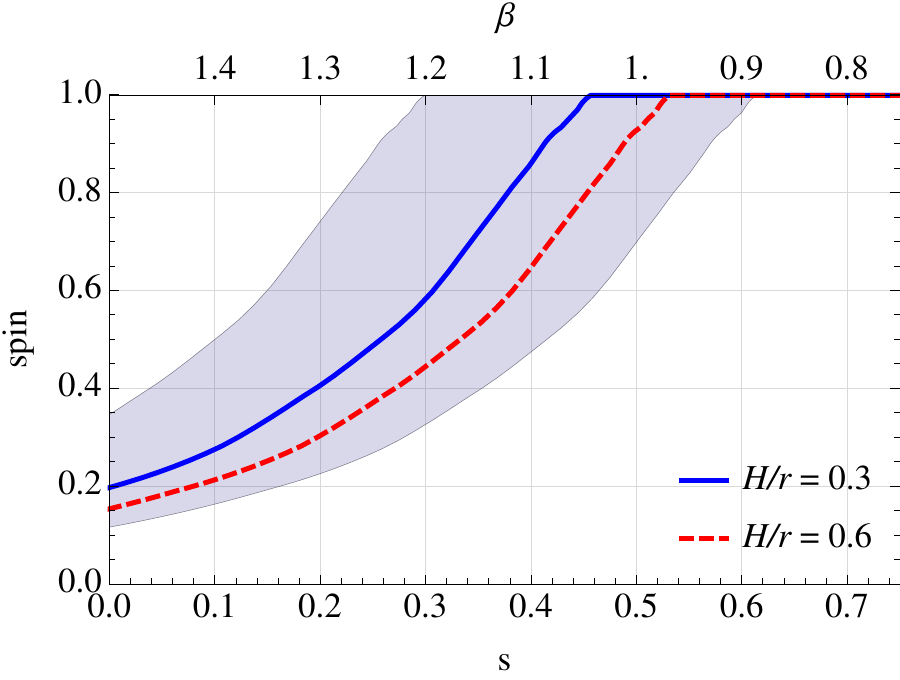}
\caption{Black hole spins that account for the jet efficiencies of the sample as a function of the mass-loss index $s$. SMBHs are assumed to be accreting in the MAD state (i.e. jet production efficiency is maximized). The jet efficiencies are estimated as in Fig. \ref{fig:median}. The thick lines represent the spins that reproduce the median efficiency of the sample for two different accretion flow thicknesses. The shaded region corresponds to the scatter around the $H/r=0.3$ model (cf. footnote 4).}
\label{fig:spins}
\end{figure}

On the assumption that the radio galaxies in our sample are accreting in the MAD state during jet production ($\eta_{\rm jet} = \eta_{\rm MAD}$) with a self-similar density profile described by the ADIOS model, we can obtain a rough estimate of the spin distribution required to explain our estimated efficiencies, as a function of the mass-loss index. Figure \ref{fig:spins} shows the spins inferred by applying the above equation to fit the median efficiency of our sample. A broad range of spins is consistent with the data. In particular, for $s>0.4$ and $H/r=0.3$, we find a $1\sigma$ lower limit of 0.5 on the spin. 

\begin{figure}
\centering
\includegraphics[width=\linewidth,trim=0 0 200 20,clip=true]{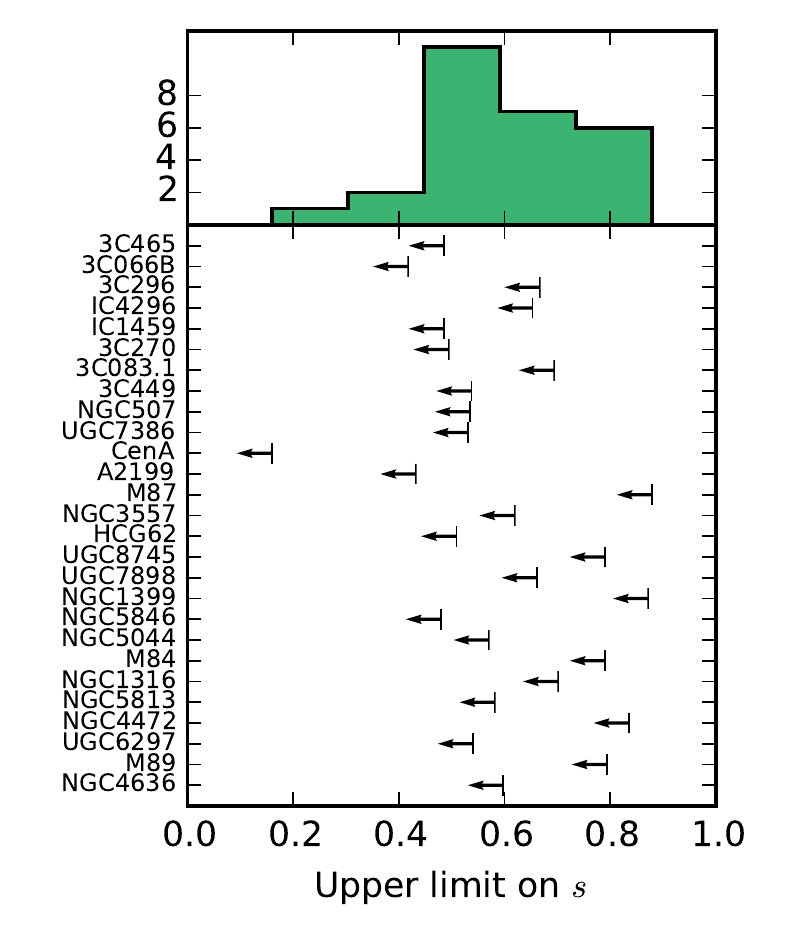}
\caption{Upper limits ($1\sigma$) on the ADIOS mass-loss index $s$, assuming that SMBHs in radio galaxies produce jets with very high efficiencies, $\eta_{\rm jet}=300\%$. The lower panel shows the individual upper limits for each AGN. The upper panel shows the histogram of limits. }
\label{fig:smax}
\end{figure}

Besides being useful for setting constraints on black hole spins, the MAD model can also be used to place upper limits on the amount of mass-loss occurring in the RIAFs. More specifically, we can obtain upper limits on $s$ for each radio galaxy by assuming that (i) during the production of powerful jets the SMBHs are accreting in the MAD state and (ii) the central engines are operating at the maximum possible kinetic efficiencies $\eta_{\rm max}$. We solve for the maximum value of $s_{\rm max}$ consistent with $\eta_{\rm max}$ for each source:
\begin{equation}
s_{\rm max} = \frac{ \log P_{\rm jet}^{\rm obs} -\log P_B^{\rm obs} -\log (\alpha \eta_{\rm max}) -1}{\log (r_i/r_B)},
\end{equation}
where we adopt the maximum jet production efficiency $\eta_{\rm max} \approx 300\%$ \citep{Sasha12,McKinney12} and $\alpha=0.1$, $r_i=10 r_S$, $r_B=10^5 r_S$ as before. We show in Fig. \ref{fig:smax} the $1\sigma$ upper limit on $s$ computed in this way for each radio galaxy.  

The median upper limit on $s$ for the sample according to Fig. \ref{fig:smax} corresponds to $\approx 0.6$ (cf. also Fig. \ref{fig:median}). From the results presented in Figures \ref{fig:spins} and \ref{fig:smax}, it follows that for $s > 0.6$, not even maximally rotating black holes accreting in the MAD state are able to account for the efficiencies of the bulk of the sample.

\subsection{Validity of jet power and Bondi inflow rate estimates}	\label{sec:bondi}

At this point, it is worth stating our basic assumptions in estimating the kinetic efficiencies, namely that:
\begin{enumerate}
\item The X-ray cavity powers are taken at face value as proxies of the jet kinetic power; more precisely, they should be regarded as the jet power averaged over the timescale during which the central engine produced one continuous -- in time -- pair of jets.
\item We can map the mass accretion rate onto the hole to the Bondi rate such that $\dot{M}_{\rm BH} \propto \dot{M}_B$.
\end{enumerate}

The first assumption is widely adopted in the literature in order to estimate the jet powers in the central radio galaxies in galaxy clusters and groups (e.g. \citealt{Cavagnolo10,OSullivan11, McNamara12,Nemmen12,Fabian12,Hlavacek-Larrondo13,King13}). Even though the estimate of the cavity power is subject to a number of uncertainties, it is perhaps the most direct way of measuring jet powers.
The assumption $E_{\rm cav}=4 PV$ corresponds to the minimum energy required to inflate the cavities and does not take into account any additional energy that may have gone into heating the intracluster medium: bubbles might be created overpressurized, such that upon injection they expand rapidly to reach pressure equilibrium with their surroundings and in the process generate weak shock waves that heat the X-ray emitting gas; there might be cosmic ray losses and even undetected cavities (cf. \citealt{McNamara12}). If such is the case, then $E_{\rm cav}$ could be a factor of a few higher than $4 PV$ (e.g., \citealt{Nusser06}).

Another possible issue with the jet power estimate is related to the estimated age of the cavity. The usual assumption of $t_{\rm age}=t_{\rm buoy}$ results in the shortest timescale by a factor of a few when compared to other possible estimates of the cavity age such as the ``terminal velocity'' timescale \citep{Birzan04,Rafferty06} or the radio lobe age obtained from synchrotron spectral analysis (e.g., \citealt{Hardcastle09,de-Gasperin12}). Despite the above-mentioned concerns with the estimates of the total jet energy and the associated timescales, it seems likely that the resulting jet powers is reasonable: the impact of underestimating the total jet energy very nearly cancels out the impact of underestimating the timescales involved (cf. discussion in \citealt{Nemmen07}). 

We should note that there are four systems in our sample that show evidence for multiple sets of cavities, which could be remnants of earlier AGN outbursts: NGC 4636 \citep{Baldi09}, NGC 5813 \citep{Randall11}, M87 \citep{Forman05} and NGC 5044 \citep{David09}. For consistency with our accretion rate estimates -- which reflect the current accreting state of the SMBHs -- the jet powers were all based on the energetics of only the innermost set of X-ray cavities for the sources \citep{Russell13}. These bubbles are still ``attached'' to the central AGN and are likely currently being inflated by the jets. Even if previous AGN outbursts were more energetic -- as seems to be the case for NGC 5813 \citep{Randall11} -- this does not impact our results which are based on inferences of the most recent AGN state. 

Regarding the second assumption, in our estimates we adopted physical models which take into account a \emph{reduction} in $\dot{M}_{\rm BH}$ with respect to $\dot{M}_B$ in the framework of the ADIOS model, assuming that the dominant source of gas supply to the black hole is the hot, diffuse, X-ray emitting gas directly observed with \emph{Chandra}. 
Nonetheless, there might be other important sources of fuel in addition to the Bondi inflow of hot gas. For instance, \cite{Ho09} estimates that the Bondi accretion of hot gas contributes roughly the same amount as stellar mass loss, $\dot{M}_* \sim \dot{M}_B$; on the other hand, \cite{Soria06b} find that stellar mass-loss can provide an amount of gas which can be an order of magnitude higher than $\dot{M}_B$ in their analysis of quiescent early-type galaxies. In principle, cold molecular gas could also play an important role in feeding the AGN (e.g., \citealt{Salome03}); however, \cite{McNamara11} found no correlation between the jet power and the total mass of cold molecular gas in a sample of brightest cluster galaxies hosting radio galaxies. In addition, turbulence and thermal instabilities may boost considerably the inflow rate at $\sim r_B$ compared to the simple estimate given by $\dot{M}_B$ \citep{Krumholz06,Gaspari13}. 

Motivated by the above observational results, we take into account the possibility of an extra contribution to the inflow rate -- e.g., potential mass loss from evolved stars within $r_B$ --  in addition to $\dot{M}_B$ which we denote as $\dot{M}_*$. We then assess the impact of such additional gas supply on the our estimates of $\eta_{\rm jet}$. We adopt for simplicity $\dot{M}_* \equiv c_* \dot{M}_B$ such that the total inflow rate supplied to the RIAF at $r_B$ is given by
\begin{equation}
\dot{M}_o=f (\dot{M}_*+\dot{M}_B),
\end{equation}
where $c_*$ is a free parameter and $f$ is a function of $\alpha$ ($f \approx 0.3$ for $\alpha=0.1$; cf. right panel of figure 4 in \citealt{Narayan11}). For example, $c_* \approx 0$ would correspond to the accretion of hot, diffuse gas from the ISM, only. We consider below three illustrative cases: (i) $c_*=1$ (amount of extra inflowing gas = amount of diffuse X-ray emitting gas); and (ii) $c_*=10$ for which the extra gas exceeds by 10x the hot inflow rate as suggested by \cite{Soria06b}.

\begin{figure}
\centering
\includegraphics[width=\linewidth]{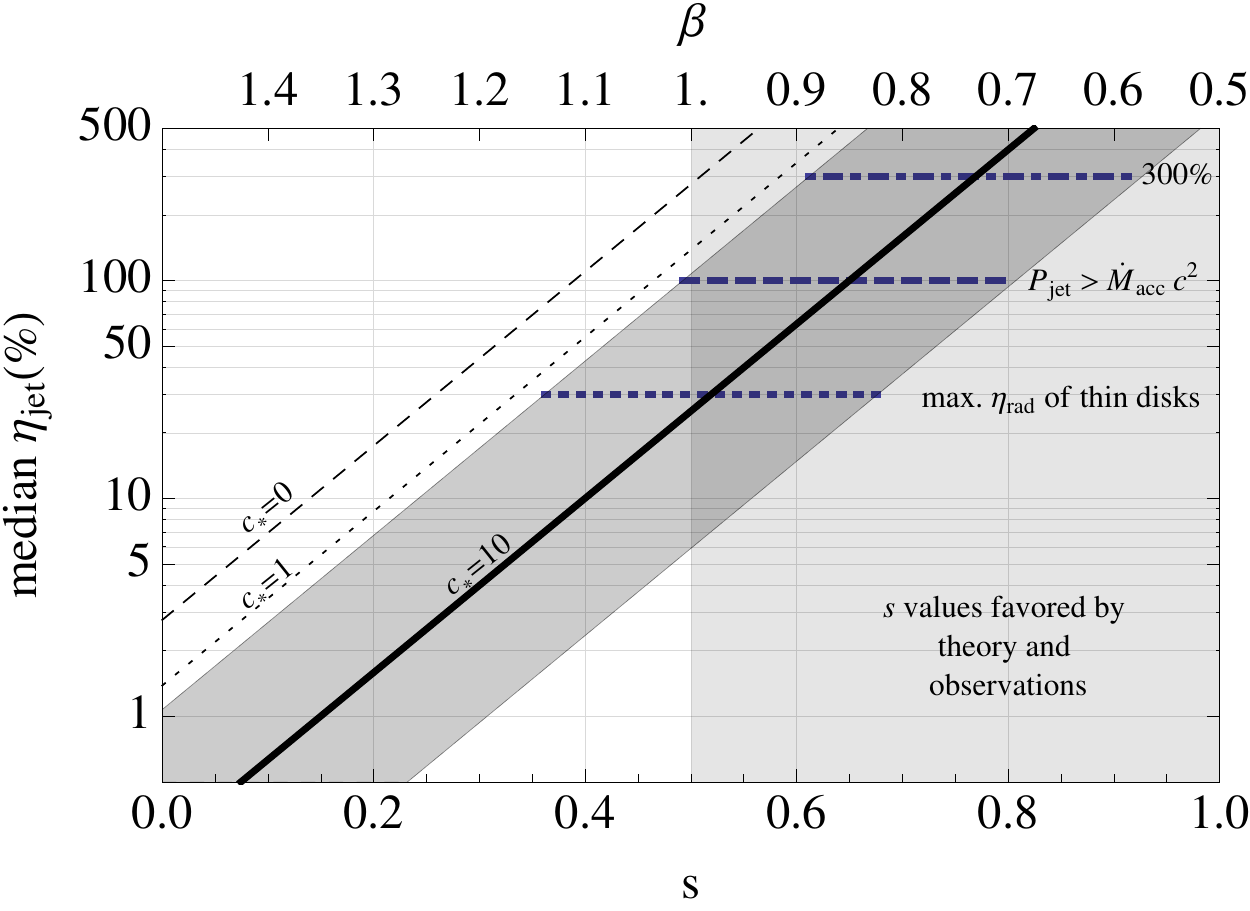}
\caption{Per cent jet kinetic efficiencies of the sample as a function of the mass-loss index $s$ for which the inflow rate at $r_B$ is increased with respect to the Bondi rate as $\dot{M}_o=f (\dot{M}_*+\dot{M}_B)$ with $\dot{M}_* \equiv c_* \dot{M}_B$. We adopt the ADIOS density profile with $\alpha=0.1$ and $r_o=10^5 r_s$. The lines represent the median efficiency of the sample computed for $c_* = 0$ (dashed), 1 (dotted) and 10 (solid). The rest of the notation follows Fig. \ref{fig:median}.}
\label{fig:moregas}
\end{figure}

Figure \ref{fig:moregas} shows the median per cent jet efficiencies estimated by using the ADIOS model for different values of $s$ and $c_*$. Considering the case of $c_*=10$, $\eta_{\rm jet}$ exceeds 30\%,100\% and 300\% for $s_{30}=0.52 \pm 0.16$, $s_{100}=0.65 \pm 0.16$ and $s_{300}=0.77 \pm 0.16$\footnote{The uncertainty on $s$ maps the sample standard deviation in $P_{\rm jet}/\dot{M}_B$.}. We conclude that more comfortable efficiencies are obtained within the range of $s$ values favored by theory and observations if there is extra inflow of gas neglected in the simple Bondi inflow estimate. 

Given that even for $c_*=0$, $\eta_{\rm jet}$ would still exceed 100\% for the bulk of the sample at $s=1$, we may conclude that an ADIOS with $s=1$ as proposed by \cite{Begelman12} is only tenable if there is a considerable extra amount of inflow (i.e. $c_* \sim 100$). 
Such high values of $c_*$ would imply that the the amount of inflowing gas due to Bondi accretion would correspond to just a perturbation compared to other sources of gas. If this was the case, we would presumably expect no $\dot{M}_B-P_{\rm jet}$ correlation, since the jet power would respond predominantly to variations in $\dot{M}_*$ which would be, in principle, decoupled from $\dot{M}_B$. The fact that we observe a correlation with a $4.2\sigma$ significance\footnote{As obtained with a simple Pearson correlation test.} argues against such high values of $c_*$.

\subsection{The scatter in $\eta_{\rm jet}$}

There are a couple of possible explanations for the significant scatter in the $\dot{M}_B c^2 - P_{\rm jet}$ relation. There is the possibility that the scatter is primarily due to the uncertainties in the determination of the cavity power and the densities at the Bondi radius, as discussed in \cite{Russell13} (see their section 4.3). Nonetheless, it is also possible that the scatter has a physical origin and there are a couple of possible physical processes that could contribute to it: 

\vspace{0.2cm}
\noindent
{\it Varying amounts of mass-loss in the sample --}
Instead of universal value of $s$ characterizing the radial accretion rate profile, there could be rather a range of $s$-values in the sample. Alternatively, the value of $s$ could remain the same in the different systems but the magnitude of mass-loss could depend on the spin such that the larger the spins, the stronger the mass-loss, as hinted by the simulation results of \cite{Sadowski13sim}.  In any case, varying amounts of mass-loss in the sample would introduce considerable scatter in the $\dot{M}_B - P_{\rm jet}$ relation, even if the jet production mechanism gives a tight $P_{\rm jet} - \dot{M}_{\rm BH}$ correlation. We find that a spread of $\sigma s \approx 0.16$ around a fixed value would be sufficient to account for the observed scatter in $P_{\rm jet}/\dot{M}_B$ (cf. Fig. \ref{fig:median}). These possibilities remain to be better explored in future simulations of accretion flows. Interestingly enough, $\sigma s$ is similar to the dispersion of $s$ values obtained in the hydrodynamic RIAF simulations of \cite{Yuan12,Bu13} for a range of initial conditions. 

\vspace{0.2cm}
\noindent
{\it Range of black hole spins and/or magnetic flux threading the horizon --}
If powerful jets are produced via the BZ mechanism then the two fundamental parameters that regulate the jet power are the black hole spin $a$ and the magnetic flux $\Phi_h$ threading the horizon, besides the mass \citep{Blandford77,Semenov04}:
\begin{equation}
P_{\rm jet} \apropto \left( \frac{a \Phi_h}{M} \right)^2;
\end{equation}
i.e., $a$ and $\Phi_h$ are degenerate to some extent (cf. \citealt{Sasha10} for refinements on the above equation). According to recent simulations, a spread in the spins at $z \sim 0$ is the naturally expected outcome of the evolution of massive black holes following their history of mergers and accretion according to models based on the hierarchical scenario of galaxy mergers \citep{Fanidakis11,Dotti13,Volonteri13}. A wide range of magnetic fluxes is also expected given the varying amounts of magnetic flux available to be accreted in the different environments of SMBHs in radio galaxies \citep{Sikora13}.

We can quantify the possible scatter in spins required to explain the scatter in the sample, by assuming that the magnetic flux around the horizon accumulated to such a point where it reached saturation (MAD state) and $H/r=0.3$. Roughly, $\sigma a \approx s$ (Fig. \ref{fig:spins}) is consistent with the scatter in spin for the different values of $s$; for $s \geq 0.5$, $a \gtrsim 0.7$ ($1\sigma$ lower limit) is required.

\vspace{0.2cm}
\noindent
{\it Variation in the amount of inflowing cold gas --}
As we discussed in section \ref{sec:bondi}, the gas input provided via mass-loss from evolved stars in a star cluster close (or inside) $r_B$ could be relevant in elliptical galaxies. In fact, a varying contribution of gas due to stellar populations which inject gas amounts comparable to $\dot{M}_B$ in the sample could also be a source of scatter in the $\dot{M}_B-P_{\rm jet}$ relation.

\section{Summary}	\label{sec:end}

With the goal of observationally constraining the jet energy efficiency $\eta_{\rm jet} \equiv P_{\rm jet}/(\dot{M}_{\rm BH} c^2)$ in radio galaxies and probing scenarios for jet production in AGNs, 
we compiled the most up-to-date dataset with constraints on the jet power and mass accretion rate onto supermassive black holes in a sample of nearby low-luminosity AGNs. Jet powers were estimated from the X-ray cavities whereas the accretion rates were obtained by combining the \emph{Chandra} X-ray constraints on the boundary conditions at the onset of the hot accretion flow (estimated within $30 r_B$ in order to minimize uncertainties involved with density extrapolations below the spatial resolution of the observatory) with current models for the density profiles of RIAFs. By studying our sample of 27 radio galaxies, we have found the following results.

(i) When correlating the jet powers with the Bondi inflow rate, we find the relation $\log P_{\rm jet} = A \log (\dot{M}_B c^2) + B$ where $A=1.1 \pm 0.2$ and $B=-4.6 \pm 8.5$ with a scatter of 0.7 dex. Our $\dot{M}_B c^2 - P_{\rm jet}$ relation is considerably less tight than found in previous studies (e.g., \citealt{Allen06}). 

(ii) If radio galaxies accrete in the ADAF mode with a viscosity parameter $\alpha \geq 0.1$, then we find the median $\eta_{\rm jet} \approx 1-3\%$ for the sample. If hot accretion flows in radio galaxies are subject to mass-loss, this efficiency estimate is likely a lower limit. 

(iii) If radio galaxies accrete in the ADIOS mode such that $\dot{M}(r) \propto r^s$, we find that the median $\eta_{\rm jet}$ of the sample exceeds 100\% for $s \geq 0.4$. This amount of mass-loss is quite on the lower end of current theoretical and observational estimates that suggest $s \gtrsim 0.5$.

(iv) The median $\eta_{\rm jet}$ exceeds the highest outflow efficiencies achieved in GRMHD simulations of jet production ($\sim 300\%$) for $s \geq 0.6$. If RIAFs in low power radio galaxies are characterized by $s \sim 0.5-0.6$, then rapidly rotating black holes $a>0.9$ and magnetically arrested disks are required in order to account for the high outflow efficiencies of the bulk of our sample. 

(iv) If RIAFs are subject to significant mass-loss, with $s$ considerably exceeding 0.6, even maximally rotating black holes and MADs are unable to provide sufficient power to the observed jets via the Blandford-Znajek process (on the assumption that the inflow rate near the accretion radius is dominated by the Bondi inflow of hot gas).

(vii) The presence of significant amounts of gas $\dot{M}_*$ inflowing at the Bondi radius in addition to the Bondi rate -- e.g., released by nuclear stellar populations -- would lower the values of $\eta_{\rm jet}$. We find that for $\dot{M}_*=10 \dot{M}_B$, the median outflow efficiencies would exceed 300\% for $s \gtrsim 0.8$.

This work highlights the fact that estimates of the outflow efficiencies of radio galaxies cannot be disentangled from the density profiles of the accretion flow. We will have to wait for the successor of \emph{Chandra} in order to obtain high spatial resolution, sub-arcsecond X-ray observations of nearby radio-loud AGNs, that resolve the sphere of influence of the massive black hole.  Such observations will allow us to constrain the density profiles similarly to what has been done for Sgr A* and NGC 3115 \citep{Wang13,Wong14}. On the theoretical side, further simulations of RIAFs with large dynamical ranges including GRMHD effects should shed light on the issue of mass-loss.

Equally important is constraining the amount of cold feeding of atomic and molecular gas versus spherical inflow of hot keV gas in powering the jets. ALMA observations of the molecular gas reservoir in radio galaxies will be instrumental in this respect. 

Very high efficiencies of jet production are quite plausible in low power radio galaxies in the context of our current understanding of RIAFs. Theoretically, efficiencies $> 100\%$ seem to be possible only when the BZ process extracts black hole spin energy most efficiently, such as when the black hole is flooded with magnetic flux (MAD state) and also rapidly rotating. The theory of MADs must therefore reach a point where clean radiative signatures of the presence of this state and spin energy extraction are developed and tested against observations of radio-loud AGNs.

\section*{Acknowledgments}

We acknowledge useful discussions with Mihoko Yukita, Ka-Wah Wong, Aleksander Sadowski, Helen Russell, Jonathan C.~McKinney, Jeremy Schnittman, Markos Georganopoulos, Ramesh Narayan, Jeff McClintock, Feng Yuan and Sylvain Guiriec. RSN was partially supported by the NASA Postdoctoral Program (NPP) at Goddard Space Flight Center, administered by Oak Ridge Associated Universities through a contract with NASA, as well as the NASA grant NNH10ZDA001N and FAPESP. AT was supported by by NASA through Einstein Postdoctoral Fellowship grant number PF3-140115 awarded by the Chandra X-ray Center, which is operated by the Smithsonian Astrophysical Observatory for NASA under contract NAS8-03060, and NASA support via High-End Computing (HEC) Program through the NASA Advanced Supercomputing (NAS) Division at Ames Research Center that provided access to the Pleiades supercomputer, as well as NSF support through an XSEDE computational time allocation TG-AST100040 on NICS Kraken, Nautilus, TACC Stampede, Maverick, and Ranch. This project made considerable use of IPython \citep{ipython} and the Astropy and mcerp libraries.

\bibliographystyle{mn2e}
\bibliography{refs}

\begin{thebibliography}{114}
\expandafter\ifx\csname natexlab\endcsname\relax\def\natexlab#1{#1}\fi

\bibitem[{{Akritas} \& {Bershady}(1996)}]{Akritas96bces}
{Akritas} M.~G., {Bershady} M.~A., 1996, \apj, 470, 706

\bibitem[{{Allen} {et~al}\mbox{.}(2006){Allen}, {Dunn}, {Fabian}, {Taylor}, \&
  {Reynolds}}]{Allen06}
{Allen} S.~W., {Dunn} R.~J.~H., {Fabian} A.~C., {Taylor} G.~B., {Reynolds}
  C.~S., 2006, \mnras, 372, 21

\bibitem[{{Baganoff} {et~al}\mbox{.}(2003){Baganoff}, {Maeda}, {Morris},
  {Bautz}, {Brandt}, {Cui}, {Doty}, {Feigelson}, {Garmire}, {Pravdo}, {Ricker},
  \& {Townsley}}]{Baganoff03}
{Baganoff} F.~K. {et~al.}, 2003, \apj, 591, 891

\bibitem[{{Baldi} {et~al}\mbox{.}(2009){Baldi}, {Forman}, {Jones}, {Kraft},
  {Nulsen}, {Churazov}, {David}, \& {Giacintucci}}]{Baldi09}
{Baldi} A., {Forman} W., {Jones} C., {Kraft} R., {Nulsen} P., {Churazov} E.,
  {David} L., {Giacintucci} S., 2009, \apj, 707, 1034

\bibitem[{{Balmaverde}, {Baldi} \& {Capetti}(2008){Balmaverde}, {Baldi}, \&
  {Capetti}}]{Balmaverde08}
{Balmaverde} B., {Baldi} R.~D., {Capetti} A., 2008, \aap, 486, 119

\bibitem[{{Begelman}(2012)}]{Begelman12}
{Begelman} M.~C., 2012, \mnras, 420, 2912

\bibitem[{Belloni {et~al}\mbox{.}(2010)Belloni {et~al.}}]{Belloni10}
Belloni T., {et~al.}, 2010, The Jet Paradigm: From Microquasars to Quasars,
  Vol. 794. Springer Verlag

\bibitem[{{B{\^\i}rzan} {et~al}\mbox{.}(2004){B{\^\i}rzan}, {Rafferty},
  {McNamara}, {Wise}, \& {Nulsen}}]{Birzan04}
{B{\^\i}rzan} L., {Rafferty} D.~A., {McNamara} B.~R., {Wise} M.~W., {Nulsen}
  P.~E.~J., 2004, \apj, 607, 800

\bibitem[{{Bisnovatyi-Kogan} \& {Ruzmaikin}(1974)}]{Bisnovatyi-Kogan74}
{Bisnovatyi-Kogan} G.~S., {Ruzmaikin} A.~A., 1974, \apss, 28, 45

\bibitem[{{Blandford} \& {Begelman}(1999)}]{Blandford99}
{Blandford} R.~D., {Begelman} M.~C., 1999, \mnras, 303, L1

\bibitem[{{Blandford} \& {Payne}(1982)}]{Blandford82}
{Blandford} R.~D., {Payne} D.~G., 1982, \mnras, 199, 883

\bibitem[{{Blandford} \& {Znajek}(1977)}]{Blandford77}
{Blandford} R.~D., {Znajek} R.~L., 1977, \mnras, 179, 433

\bibitem[{{Bondi}(1952)}]{Bondi52}
{Bondi} H., 1952, \mnras, 112, 195

\bibitem[{{Bu} {et~al}\mbox{.}(2013){Bu}, {Yuan}, {Wu}, \& {Cuadra}}]{Bu13}
{Bu} D.-F., {Yuan} F., {Wu} M., {Cuadra} J., 2013, \mnras, 434, 1692

\bibitem[{{Cavagnolo} {et~al}\mbox{.}(2010){Cavagnolo}, {McNamara}, {Nulsen},
  {Carilli}, {Jones}, \& {B{\^\i}rzan}}]{Cavagnolo10}
{Cavagnolo} K.~W., {McNamara} B.~R., {Nulsen} P.~E.~J., {Carilli} C.~L.,
  {Jones} C., {B{\^\i}rzan} L., 2010, \apj, 720, 1066

\bibitem[{{Churazov} {et~al}\mbox{.}(2002){Churazov}, {Sunyaev}, {Forman}, \&
  {B{\"o}hringer}}]{Churazov02}
{Churazov} E., {Sunyaev} R., {Forman} W., {B{\"o}hringer} H., 2002, \mnras,
  332, 729

\bibitem[{{Daly} \& {Sprinkle}(2014)}]{Daly14}
{Daly} R.~A., {Sprinkle} T.~B., 2014, \mnras, 438, 3233

\bibitem[{{David} {et~al}\mbox{.}(2009){David}, {Jones}, {Forman}, {Nulsen},
  {Vrtilek}, {O'Sullivan}, {Giacintucci}, \& {Raychaudhury}}]{David09}
{David} L.~P., {Jones} C., {Forman} W., {Nulsen} P., {Vrtilek} J., {O'Sullivan}
  E., {Giacintucci} S., {Raychaudhury} S., 2009, \apj, 705, 624

\bibitem[{{de Gasperin} {et~al}\mbox{.}(2012){de Gasperin}, {Orr{\'u}},
  {Murgia}, {Merloni}, {Falcke}, {Beck}, {Beswick}, {B{\^\i}rzan}, {Bonafede},
  {Br{\"u}ggen}, {Brunetti}, {Chy{\.z}y}, {Conway}, {Croston}, {En{\ss}lin},
  {Ferrari}, {Heald}, {Heidenreich}, {Jackson}, {Macario}, {McKean}, {Miley},
  {Morganti}, {Offringa}, {Pizzo}, {Rafferty}, {R{\"o}ttgering}, {Shulevski},
  {Steinmetz}, {Tasse}, {van der Tol}, {van Driel}, {van Weeren}, {van
  Zwieten}, {Alexov}, {Anderson}, {Asgekar}, {Avruch}, {Bell}, {Bell},
  {Bentum}, {Bernardi}, {Best}, {Breitling}, {Broderick}, {Butcher}, {Ciardi},
  {Dettmar}, {Eisloeffel}, {Frieswijk}, {Gankema}, {Garrett}, {Gerbers},
  {Griessmeier}, {Gunst}, {Hassall}, {Hessels}, {Hoeft}, {Horneffer},
  {Karastergiou}, {K{\"o}hler}, {Koopman}, {Kuniyoshi}, {Kuper}, {Maat},
  {Mann}, {Mevius}, {Mulcahy}, {Munk}, {Nijboer}, {Noordam}, {Paas}, {Pandey},
  {Pandey}, {Polatidis}, {Reich}, {Schoenmakers}, {Sluman}, {Smirnov}, {Sobey},
  {Stappers}, {Swinbank}, {Tagger}, {Tang}, {van Bemmel}, {van Cappellen}, {van
  Duin}, {van Haarlem}, {van Leeuwen}, {Vermeulen}, {Vocks}, {White}, {Wise},
  {Wucknitz}, \& {Zarka}}]{de-Gasperin12}
{de Gasperin} F. {et~al.}, 2012, \aap, 547, A56

\bibitem[{{De Villiers}, {Hawley} \& {Krolik}(2003){De Villiers}, {Hawley}, \&
  {Krolik}}]{De-Villiers03}
{De Villiers} J.-P., {Hawley} J.~F., {Krolik} J.~H., 2003, \apj, 599, 1238

\bibitem[{{Di Matteo} {et~al}\mbox{.}(2003){Di Matteo}, {Allen}, {Fabian},
  {Wilson}, \& {Young}}]{Di-Matteo03}
{Di Matteo} T., {Allen} S.~W., {Fabian} A.~C., {Wilson} A.~S., {Young} A.~J.,
  2003, \apj, 582, 133

\bibitem[{{Dotti} {et~al}\mbox{.}(2013){Dotti}, {Colpi}, {Pallini}, {Perego},
  \& {Volonteri}}]{Dotti13}
{Dotti} M., {Colpi} M., {Pallini} S., {Perego} A., {Volonteri} M., 2013, \apj,
  762, 68

\bibitem[{{Dunn} \& {Fabian}(2004)}]{Dunn04}
{Dunn} R.~J.~H., {Fabian} A.~C., 2004, \mnras, 355, 862

\bibitem[{{Dunn}, {Fabian} \& {Taylor}(2005){Dunn}, {Fabian}, \&
  {Taylor}}]{Dunn05}
{Dunn} R.~J.~H., {Fabian} A.~C., {Taylor} G.~B., 2005, \mnras, 364, 1343

\bibitem[{{Fabian}(2012)}]{Fabian12}
{Fabian} A.~C., 2012, \araa, 50, 455

\bibitem[{{Fanidakis} {et~al}\mbox{.}(2011){Fanidakis}, {Baugh}, {Benson},
  {Bower}, {Cole}, {Done}, \& {Frenk}}]{Fanidakis11}
{Fanidakis} N., {Baugh} C.~M., {Benson} A.~J., {Bower} R.~G., {Cole} S., {Done}
  C., {Frenk} C.~S., 2011, \mnras, 410, 53

\bibitem[{{Fender} \& {Belloni}(2012)}]{Fender12}
{Fender} R., {Belloni} T., 2012, Science, 337, 540

\bibitem[{{Fender}, {Gallo} \& {Russell}(2010){Fender}, {Gallo}, \&
  {Russell}}]{Fender10}
{Fender} R.~P., {Gallo} E., {Russell} D., 2010, \mnras, 406, 1425

\bibitem[{{Forman} {et~al}\mbox{.}(2005){Forman}, {Nulsen}, {Heinz}, {Owen},
  {Eilek}, {Vikhlinin}, {Markevitch}, {Kraft}, {Churazov}, \&
  {Jones}}]{Forman05}
{Forman} W. {et~al.}, 2005, \apj, 635, 894

\bibitem[{{Fragile}, {Wilson} \& {Rodriguez}(2012){Fragile}, {Wilson}, \&
  {Rodriguez}}]{Fragile12jet}
{Fragile} P.~C., {Wilson} J., {Rodriguez} M., 2012, \mnras, 424, 524

\bibitem[{{Gardner} \& {Done}(2014)}]{Gardner14}
{Gardner} E., {Done} C., 2014, \mnras, 438, 779

\bibitem[{{Gaspari}, {Ruszkowski} \& {Oh}(2013){Gaspari}, {Ruszkowski}, \&
  {Oh}}]{Gaspari13}
{Gaspari} M., {Ruszkowski} M., {Oh} S.~P., 2013, \mnras, 432, 3401

\bibitem[{{Ghisellini} {et~al}\mbox{.}(2014){Ghisellini}, {Tavecchio},
  {Maraschi}, {Celotti}, \& {Sbarrato}}]{Ghisellini14}
{Ghisellini} G., {Tavecchio} F., {Maraschi} L., {Celotti} A., {Sbarrato} T.,
  2014, \nat, 515, 376

\bibitem[{{Ghosh} \& {Abramowicz}(1997)}]{Ghosh97}
{Ghosh} P., {Abramowicz} M.~A., 1997, \mnras, 292, 887

\bibitem[{{Hardcastle} {et~al}\mbox{.}(2009){Hardcastle}, {Cheung}, {Feain}, \&
  {Stawarz}}]{Hardcastle09}
{Hardcastle} M.~J., {Cheung} C.~C., {Feain} I.~J., {Stawarz} {\L}., 2009,
  \mnras, 393, 1041

\bibitem[{{Hawley}, {Guan} \& {Krolik}(2011){Hawley}, {Guan}, \&
  {Krolik}}]{Hawley11}
{Hawley} J.~F., {Guan} X., {Krolik} J.~H., 2011, \apj, 738, 84

\bibitem[{{Hawley} \& {Krolik}(2001)}]{Hawley01}
{Hawley} J.~F., {Krolik} J.~H., 2001, \apj, 548, 348

\bibitem[{{Hawley} \& {Krolik}(2006)}]{Hawley06}
{Hawley} J.~F., {Krolik} J.~H., 2006, \apj, 641, 103

\bibitem[{{Heinz}, {Merloni} \& {Schwab}(2007){Heinz}, {Merloni}, \&
  {Schwab}}]{Heinz07}
{Heinz} S., {Merloni} A., {Schwab} J., 2007, \apjl, 658, L9

\bibitem[{{Hlavacek-Larrondo} {et~al}\mbox{.}(2013){Hlavacek-Larrondo},
  {Allen}, {Taylor}, {Fabian}, {Canning}, {Werner}, {Sanders}, {Grimes},
  {Ehlert}, \& {von der Linden}}]{Hlavacek-Larrondo13}
{Hlavacek-Larrondo} J. {et~al.}, 2013, \apj, 777, 163

\bibitem[{{Ho}(2009)}]{Ho09}
{Ho} L.~C., 2009, \apj, 699, 626

\bibitem[{{Hughes} \& {Blandford}(2003)}]{Hughes03}
{Hughes} S.~A., {Blandford} R.~D., 2003, \apjl, 585, L101

\bibitem[{{Igumenshchev}, {Narayan} \& {Abramowicz}(2003){Igumenshchev},
  {Narayan}, \& {Abramowicz}}]{Igumenshchev03}
{Igumenshchev} I.~V., {Narayan} R., {Abramowicz} M.~A., 2003, \apj, 592, 1042

\bibitem[{{King} {et~al}\mbox{.}(2013{\natexlab{a}}){King}, {Miller},
  {G{\"u}ltekin}, {Walton}, {Fabian}, {Reynolds}, \& {Nandra}}]{King13spin}
{King} A.~L., {Miller} J.~M., {G{\"u}ltekin} K., {Walton} D.~J., {Fabian}
  A.~C., {Reynolds} C.~S., {Nandra} K., 2013{\natexlab{a}}, \apj, 771, 84

\bibitem[{{King} {et~al}\mbox{.}(2013{\natexlab{b}}){King}, {Miller},
  {Raymond}, {Fabian}, {Reynolds}, {G{\"u}ltekin}, {Cackett}, {Allen}, {Proga},
  \& {Kallman}}]{King13}
{King} A.~L. {et~al.}, 2013{\natexlab{b}}, \apj, 762, 103

\bibitem[{{Koide} {et~al}\mbox{.}(2002){Koide}, {Shibata}, {Kudoh}, \&
  {Meier}}]{Koide02}
{Koide} S., {Shibata} K., {Kudoh} T., {Meier} D.~L., 2002, Science, 295, 1688

\bibitem[{{Komissarov}(2001)}]{Komissarov01}
{Komissarov} S.~S., 2001, \mnras, 326, L41

\bibitem[{{Komissarov}(2005)}]{2005MNRAS.359..801K}
{Komissarov} S.~S., 2005, \mnras, 359, 801

\bibitem[{{Krumholz}, {McKee} \& {Klein}(2006){Krumholz}, {McKee}, \&
  {Klein}}]{Krumholz06}
{Krumholz} M.~R., {McKee} C.~F., {Klein} R.~I., 2006, \apj, 638, 369

\bibitem[{{Kuo} {et~al}\mbox{.}(2014){Kuo}, {Asada}, {Rao}, {Nakamura},
  {Algaba}, {Liu}, {Inoue}, {Koch}, {Ho}, {Matsushita}, {Pu}, {Akiyama},
  {Nishioka}, \& {Pradel}}]{Kuo14}
{Kuo} C.~Y. {et~al.}, 2014, \apjl, 783, L33

\bibitem[{{Li}, {Ostriker} \& {Sunyaev}(2013){Li}, {Ostriker}, \&
  {Sunyaev}}]{Li13}
{Li} J., {Ostriker} J., {Sunyaev} R., 2013, \apj, 767, 105

\bibitem[{{Livio}, {Ogilvie} \& {Pringle}(1999){Livio}, {Ogilvie}, \&
  {Pringle}}]{Livio99}
{Livio} M., {Ogilvie} G.~I., {Pringle} J.~E., 1999, \apj, 512, 100

\bibitem[{{Marrone} {et~al}\mbox{.}(2007){Marrone}, {Moran}, {Zhao}, \&
  {Rao}}]{Marrone07}
{Marrone} D.~P., {Moran} J.~M., {Zhao} J.-H., {Rao} R., 2007, \apjl, 654, L57

\bibitem[{{Mart{\'{\i}}nez-Sansigre} \& {Rawlings}(2011)}]{Martinez-Sansigre11}
{Mart{\'{\i}}nez-Sansigre} A., {Rawlings} S., 2011, \mnras, 414, 1937

\bibitem[{{McKinney}(2005)}]{McKinney05}
{McKinney} J.~C., 2005, \apjl, 630, L5

\bibitem[{{McKinney} \& {Gammie}(2004)}]{McKinney04}
{McKinney} J.~C., {Gammie} C.~F., 2004, \apj, 611, 977

\bibitem[{{McKinney}, {Tchekhovskoy} \& {Blandford}(2012){McKinney},
  {Tchekhovskoy}, \& {Blandford}}]{McKinney12}
{McKinney} J.~C., {Tchekhovskoy} A., {Blandford} R.~D., 2012, \mnras, 423, 3083

\bibitem[{{McNamara} \& {Nulsen}(2007)}]{McNamara07}
{McNamara} B.~R., {Nulsen} P.~E.~J., 2007, \araa, 45, 117

\bibitem[{{McNamara} \& {Nulsen}(2012)}]{McNamara12}
{McNamara} B.~R., {Nulsen} P.~E.~J., 2012, New Journal of Physics, 14, 055023

\bibitem[{{McNamara}, {Rohanizadegan} \& {Nulsen}(2011){McNamara},
  {Rohanizadegan}, \& {Nulsen}}]{McNamara11}
{McNamara} B.~R., {Rohanizadegan} M., {Nulsen} P.~E.~J., 2011, \apj, 727, 39

\bibitem[{{Meier}(2012)}]{Meier12}
{Meier} D.~L., 2012, {Black Hole Astrophysics: The Engine Paradigm}. Springer,
  Verlag Berlin Heidelberg

\bibitem[{{Merloni} \& {Heinz}(2007)}]{Merloni07}
{Merloni} A., {Heinz} S., 2007, \mnras, 381, 589

\bibitem[{{Narayan}(2002)}]{Narayan02}
{Narayan} R., 2002, in Lighthouses of the Universe: The Most Luminous Celestial
  Objects and Their Use for Cosmology, {Gilfanov} M., {Sunyeav} R., {Churazov}
  E., eds., p. 405

\bibitem[{{Narayan} \& {Fabian}(2011)}]{Narayan11}
{Narayan} R., {Fabian} A.~C., 2011, \mnras, 415, 3721

\bibitem[{{Narayan}, {Igumenshchev} \& {Abramowicz}(2003){Narayan},
  {Igumenshchev}, \& {Abramowicz}}]{Narayan03}
{Narayan} R., {Igumenshchev} I.~V., {Abramowicz} M.~A., 2003, \pasj, 55, L69

\bibitem[{{Narayan} \& {McClintock}(2008)}]{Narayan08}
{Narayan} R., {McClintock} J.~E., 2008, New Astron. Rev., 51, 733

\bibitem[{{Narayan} \& {McClintock}(2012)}]{Narayan12}
{Narayan} R., {McClintock} J.~E., 2012, \mnras, 419, L69

\bibitem[{{Narayan} {et~al}\mbox{.}(2012){Narayan}, {Sadowski}, {Penna}, \&
  {Kulkarni}}]{Narayan12grmhd}
{Narayan} R., {Sadowski} A., {Penna} R.~F., {Kulkarni} A.~K., 2012, \mnras,
  426, 3241

\bibitem[{{Narayan} \& {Yi}(1994)}]{Narayan94}
{Narayan} R., {Yi} I., 1994, \apjl, 428, L13

\bibitem[{{Narayan} \& {Yi}(1995)}]{Narayan95b}
{Narayan} R., {Yi} I., 1995, \apj, 444, 231

\bibitem[{{Nemmen} {et~al}\mbox{.}(2007){Nemmen}, {Bower}, {Babul}, \&
  {Storchi-Bergmann}}]{Nemmen07}
{Nemmen} R.~S., {Bower} R.~G., {Babul} A., {Storchi-Bergmann} T., 2007, \mnras,
  377, 1652

\bibitem[{{Nemmen} {et~al}\mbox{.}(2012){Nemmen}, {Georganopoulos}, {Guiriec},
  {Meyer}, {Gehrels}, \& {Sambruna}}]{Nemmen12}
{Nemmen} R.~S., {Georganopoulos} M., {Guiriec} S., {Meyer} E.~T., {Gehrels} N.,
  {Sambruna} R.~M., 2012, Science, 338, 1445

\bibitem[{{Nemmen}, {Storchi-Bergmann} \& {Eracleous}(2014){Nemmen},
  {Storchi-Bergmann}, \& {Eracleous}}]{Nemmen14}
{Nemmen} R.~S., {Storchi-Bergmann} T., {Eracleous} M., 2014, \mnras, 438, 2804

\bibitem[{{Nusser}, {Silk} \& {Babul}(2006){Nusser}, {Silk}, \&
  {Babul}}]{Nusser06}
{Nusser} A., {Silk} J., {Babul} A., 2006, \mnras, 373, 739

\bibitem[{{O'Sullivan} {et~al}\mbox{.}(2011){O'Sullivan}, {Giacintucci},
  {David}, {Gitti}, {Vrtilek}, {Raychaudhury}, \& {Ponman}}]{OSullivan11}
{O'Sullivan} E., {Giacintucci} S., {David} L.~P., {Gitti} M., {Vrtilek} J.~M.,
  {Raychaudhury} S., {Ponman} T.~J., 2011, \apj, 735, 11

\bibitem[{{Pang} {et~al}\mbox{.}(2011){Pang}, {Pen}, {Matzner}, {Green}, \&
  {Liebend{\"o}rfer}}]{Pang11}
{Pang} B., {Pen} U.-L., {Matzner} C.~D., {Green} S.~R., {Liebend{\"o}rfer} M.,
  2011, \mnras, 415, 1228

\bibitem[{{Penna} {et~al}\mbox{.}(2013){Penna}, {S{\c a}dowski}, {Kulkarni}, \&
  {Narayan}}]{Penna13}
{Penna} R.~F., {S{\c a}dowski} A., {Kulkarni} A.~K., {Narayan} R., 2013,
  \mnras, 428, 2255

\bibitem[{P\'erez \& Granger(2007)}]{ipython}
P\'erez F., Granger B.~E., 2007, {C}omput. {S}ci. {E}ng., 9, 21

\bibitem[{{Proga} \& {Begelman}(2003)}]{Proga03}
{Proga} D., {Begelman} M.~C., 2003, \apj, 592, 767

\bibitem[{{Rafferty} {et~al}\mbox{.}(2006){Rafferty}, {McNamara}, {Nulsen}, \&
  {Wise}}]{Rafferty06}
{Rafferty} D.~A., {McNamara} B.~R., {Nulsen} P.~E.~J., {Wise} M.~W., 2006,
  \apj, 652, 216

\bibitem[{{Randall} {et~al}\mbox{.}(2011){Randall}, {Forman}, {Giacintucci},
  {Nulsen}, {Sun}, {Jones}, {Churazov}, {David}, {Kraft}, {Donahue}, {Blanton},
  {Simionescu}, \& {Werner}}]{Randall11}
{Randall} S.~W. {et~al.}, 2011, \apj, 726, 86

\bibitem[{{Reynolds}(2013)}]{Reynolds13}
{Reynolds} C.~S., 2013, Classical and Quantum Gravity, 30, 244004

\bibitem[{{Russell}, {Gallo} \& {Fender}(2013){Russell}, {Gallo}, \&
  {Fender}}]{Russell13bhb}
{Russell} D.~M., {Gallo} E., {Fender} R.~P., 2013, \mnras, 431, 405

\bibitem[{{Russell} {et~al}\mbox{.}(2013){Russell}, {McNamara}, {Edge},
  {Hogan}, {Main}, \& {Vantyghem}}]{Russell13}
{Russell} H.~R., {McNamara} B.~R., {Edge} A.~C., {Hogan} M.~T., {Main} R.~A.,
  {Vantyghem} A.~N., 2013, \mnras, 432, 530

\bibitem[{{Salom{\'e}} \& {Combes}(2003)}]{Salome03}
{Salom{\'e}} P., {Combes} F., 2003, \aap, 412, 657

\bibitem[{{S{\c a}dowski} {et~al}\mbox{.}(2013){S{\c a}dowski}, {Narayan},
  {Penna}, \& {Zhu}}]{Sadowski13sim}
{S{\c a}dowski} A., {Narayan} R., {Penna} R., {Zhu} Y., 2013, \mnras, 436, 3856

\bibitem[{{Semenov}, {Dyadechkin} \& {Punsly}(2004){Semenov}, {Dyadechkin}, \&
  {Punsly}}]{Semenov04}
{Semenov} V., {Dyadechkin} S., {Punsly} B., 2004, Science, 305, 978

\bibitem[{{Shakura} \& {Sunyaev}(1973)}]{Shakura73}
{Shakura} N.~I., {Sunyaev} R.~A., 1973, \aap, 24, 337

\bibitem[{{Sif{\'o}n} {et~al}\mbox{.}(2013){Sif{\'o}n}, {Menanteau},
  {Hasselfield}, {Marriage}, {Hughes}, {Barrientos}, {Gonz{\'a}lez}, {Infante},
  {Addison}, {Baker}, {Battaglia}, {Bond}, {Crichton}, {Das}, {Devlin},
  {Dunkley}, {D{\"u}nner}, {Gralla}, {Hajian}, {Hilton}, {Hincks}, {Kosowsky},
  {Marsden}, {Moodley}, {Niemack}, {Nolta}, {Page}, {Partridge}, {Reese},
  {Sehgal}, {Sievers}, {Spergel}, {Staggs}, {Thornton}, {Trac}, \&
  {Wollack}}]{Sifon13}
{Sif{\'o}n} C. {et~al.}, 2013, \apj, 772, 25

\bibitem[{{Sikora} \& {Begelman}(2013)}]{Sikora13}
{Sikora} M., {Begelman} M.~C., 2013, \apjl, 764, L24

\bibitem[{{Sikora}, {Stawarz} \& {Lasota}(2007){Sikora}, {Stawarz}, \&
  {Lasota}}]{Sikora07}
{Sikora} M., {Stawarz} {\L}., {Lasota} J.-P., 2007, \apj, 658, 815

\bibitem[{{Soria} {et~al}\mbox{.}(2006){Soria}, {Graham}, {Fabbiano}, {Baldi},
  {Elvis}, {Jerjen}, {Pellegrini}, \& {Siemiginowska}}]{Soria06b}
{Soria} R., {Graham} A.~W., {Fabbiano} G., {Baldi} A., {Elvis} M., {Jerjen} H.,
  {Pellegrini} S., {Siemiginowska} A., 2006, \apj, 640, 143

\bibitem[{{Steiner}, {McClintock} \& {Narayan}(2013){Steiner}, {McClintock}, \&
  {Narayan}}]{Steiner13}
{Steiner} J.~F., {McClintock} J.~E., {Narayan} R., 2013, \apj, 762, 104

\bibitem[{{Stone}, {Pringle} \& {Begelman}(1999){Stone}, {Pringle}, \&
  {Begelman}}]{Stone99}
{Stone} J.~M., {Pringle} J.~E., {Begelman} M.~C., 1999, \mnras, 310, 1002

\bibitem[{{Tchekhovskoy}(2015)}]{sasha15}
{Tchekhovskoy} A., 2015, in Astrophysics and Space Science Library, Vol. 414,
  Astrophysics and Space Science Library, {Contopoulos} I., {Gabuzda} D.,
  {Kylafis} N., eds., p.~45

\bibitem[{{Tchekhovskoy} \& {McKinney}(2012)}]{Sasha12}
{Tchekhovskoy} A., {McKinney} J.~C., 2012, \mnras, 423, L55

\bibitem[{{Tchekhovskoy}, {McKinney} \& {Narayan}(2012){Tchekhovskoy},
  {McKinney}, \& {Narayan}}]{Sasha12rev}
{Tchekhovskoy} A., {McKinney} J.~C., {Narayan} R., 2012, Journal of Physics
  Conference Series, 372, 012040

\bibitem[{{Tchekhovskoy}, {Narayan} \& {McKinney}(2010){Tchekhovskoy},
  {Narayan}, \& {McKinney}}]{Sasha10}
{Tchekhovskoy} A., {Narayan} R., {McKinney} J.~C., 2010, \apj, 711, 50

\bibitem[{{Tchekhovskoy}, {Narayan} \& {McKinney}(2011){Tchekhovskoy},
  {Narayan}, \& {McKinney}}]{Sasha11}
{Tchekhovskoy} A., {Narayan} R., {McKinney} J.~C., 2011, \mnras, 418, L79

\bibitem[{{Thorne}(1974)}]{Thorne74}
{Thorne} K.~S., 1974, \apj, 191, 507

\bibitem[{{van Velzen} \& {Falcke}(2013)}]{van-Velzen13}
{van Velzen} S., {Falcke} H., 2013, \aap, 557, L7

\bibitem[{{Vattakunnel} {et~al}\mbox{.}(2010){Vattakunnel}, {Trussoni},
  {Capetti}, \& {Baldi}}]{Vattakunnel10}
{Vattakunnel} S., {Trussoni} E., {Capetti} A., {Baldi} R.~D., 2010, \aap, 522,
  A89

\bibitem[{{Volonteri}(2012)}]{Volonteri12}
{Volonteri} M., 2012, Science, 337, 544

\bibitem[{{Volonteri}, {Sikora} \& {Lasota}(2007){Volonteri}, {Sikora}, \&
  {Lasota}}]{Volonteri07}
{Volonteri} M., {Sikora} M., {Lasota} J.-P., 2007, \apj, 667, 704

\bibitem[{{Volonteri} {et~al}\mbox{.}(2013){Volonteri}, {Sikora}, {Lasota}, \&
  {Merloni}}]{Volonteri13}
{Volonteri} M., {Sikora} M., {Lasota} J.-P., {Merloni} A., 2013, \apj, 775, 94

\bibitem[{Wang {et~al}\mbox{.}(2013)Wang, Nowak, Markoff, Baganoff, Nayakshin,
  Yuan, Cuadra, Davis, Dexter, Fabian, Grosso, Haggard, Houck, Ji, Li, Neilsen,
  Porquet, Ripple, \& Shcherbakov}]{Wang13}
Wang Q.~D. {et~al.}, 2013, Science, 341, 981

\bibitem[{{Wong} {et~al}\mbox{.}(2014){Wong}, {Irwin}, {Shcherbakov}, {Yukita},
  {Million}, \& {Bregman}}]{Wong14}
{Wong} K.-W., {Irwin} J.~A., {Shcherbakov} R.~V., {Yukita} M., {Million} E.~T.,
  {Bregman} J.~N., 2014, \apj, 780, 9

\bibitem[{{Wong} {et~al}\mbox{.}(2011){Wong}, {Irwin}, {Yukita}, {Million},
  {Mathews}, \& {Bregman}}]{Wong11}
{Wong} K.-W., {Irwin} J.~A., {Yukita} M., {Million} E.~T., {Mathews} W.~G.,
  {Bregman} J.~N., 2011, \apjl, 736, L23

\bibitem[{{Wu}, {Yuan} \& {Cao}(2007){Wu}, {Yuan}, \& {Cao}}]{Wu07}
{Wu} Q., {Yuan} F., {Cao} X., 2007, \apj, 669, 96

\bibitem[{{Yuan}, {Bu} \& {Wu}(2012){Yuan}, {Bu}, \& {Wu}}]{Yuan12b}
{Yuan} F., {Bu} D., {Wu} M., 2012, \apj, 761, 130

\bibitem[{{Yuan} \& {Narayan}(2014)}]{Yuan14}
{Yuan} F., {Narayan} R., 2014, \araa, 52, 529

\bibitem[{{Yuan}, {Quataert} \& {Narayan}(2003){Yuan}, {Quataert}, \&
  {Narayan}}]{Yuan03}
{Yuan} F., {Quataert} E., {Narayan} R., 2003, \apj, 598, 301

\bibitem[{{Yuan}, {Wu} \& {Bu}(2012){Yuan}, {Wu}, \& {Bu}}]{Yuan12}
{Yuan} F., {Wu} M., {Bu} D., 2012, \apj, 761, 129

\bibitem[{Zamaninasab {et~al}\mbox{.}(2014)Zamaninasab, Clausen-Brown,
  Savolainen, \& Tchekhovskoy}]{Zamaninasab14}
Zamaninasab M., Clausen-Brown E., Savolainen T., Tchekhovskoy A., 2014, Nature,
  510, 126

\end{thebibliography}

\bsp

\label{lastpage}

\end{document}